\documentclass[12pt,draftcls,onecolumn]{IEEEtran}
\usepackage{graphicx}
\usepackage{color}
\usepackage{fancyhdr}
\usepackage[cmex10]{amsmath}
\usepackage{mdwmath}
\usepackage{amssymb}
\usepackage{textcomp}
\usepackage{times}
\usepackage{multicol}
\usepackage{float}
\usepackage[numbers,sort&compress]{natbib}
\usepackage{color, soul}
\usepackage{subfig}

\usepackage{color}
\usepackage{algorithm} 
\usepackage{algorithmic} 
\usepackage{multirow}
\usepackage{amsmath,bm}

\ifCLASSINFOpdf
\else
\fi
\begin{document}
\title{Position-aided Large-scale MIMO Channel Estimation for High-Speed Railway Communication Systems}




\author{Tao~Li,~Xiaodong~Wang,~\IEEEmembership{Fellow,~IEEE},~Pingyi~Fan,~\IEEEmembership{Senior~Member,~IEEE,} and Taneli Riihonen,~\IEEEmembership{Member,~IEEE}
\thanks{T. Li and P. Fan are with Tsinghua National Laboratory for Information Science and Technology (TNList) as well as Department of EE, Tsinghua University, Beijing, P. R. China, 100084 (e-mail: litao12@mails.tsinghua.edu.cn; fpy@tsinghua.edu.cn).

X. Wang is with the Electrical Engineering Department, Columbia University, New York, NY 10027, USA (e-mail: wangx@ee.columbia.edu).

T. Riihonen is with the Department of Signal Processing and Acoustics, Aalto University School of Electrical
Engineering, 00076 Aalto, Finland (e-mail: taneli.riihonen@aalto.fi).}}

\maketitle
\thispagestyle{empty}
\baselineskip 24pt
\begin{abstract}
\baselineskip 18pt
We consider channel estimation for high-speed railway communication systems, where both the transmitter and the receiver are equipped with large-scale antenna arrays. It is known that the throughput of conventional training schemes monotonically decreases with the mobility.
Assuming that the moving terminal employs a large linear antenna array, this paper proposes a position-aided channel estimation scheme whereby only a portion of the transmit antennas send pilot symbols and the full channel matrix can be well estimated by using these pilots together with the antenna position information based on the joint spatial-temporal correlation. The relationship between mobility and throughput/DoF is established. Furthermore, the optimal selections of transmit power and time interval partition between the training and data phases as well as the antenna size are presented accordingly. Both analytical and simulation results show that the system throughput with the position-aided channel estimator does not deteriorate appreciably as the mobility increases, which is sharply in contrast with the conventional one.
\end{abstract}

\begin{IEEEkeywords}
  Large-scale \mbox{MIMO}, high-speed railway communications, linear antenna array, channel estimation, joint spatio-temporal correlation, throughput, degree of freedom (DoF).
\end{IEEEkeywords}

%
\IEEEpeerreviewmaketitle

\section{Introduction}
%
%
%
%
The large-scale multiple-input multiple-output (MIMO) technology holds the key to significantly improving the throughput of future wireless communication systems \cite{Goldsmith_4}. For high-speed railway communication systems, both the base station (BS) and the mobile terminal (i.e., the train) can employ large-scale antenna arrays to provide high-throughput services to users on the train \cite{Luo_4,Wang_4,Ghazal_18}. In this paper, we focus on such a high-speed railway MIMO scenario, where both the transmitter and the receiver are equipped with large-scale antenna arrays.

As we know, in MIMO communications, to obtain the instantaneous channel state information (CSI), the training-based channel estimator is widely used. Although the training overhead may be insignificant in single-antenna systems, it becomes the major impediment to high-speed railway MIMO communications, where the speed of the mobile terminal can reach up to hundreds of kilometers per hour \cite{Pan_32}. In particular, the throughput of the large-scale MIMO system even can deteriorate to zero if the training phase occupies all the channel uses \cite{Hassibi_5}. It seems very pessimistic to employ large-scale MIMO in highly mobile environments, because the high time-selectivity of the channel removes the benefits brought by multi-antenna wireless links \cite{Larsson_3, Heath_6}.

A rich body of the research in the literature focused on the training-based channel estimation for large-scale MIMO systems under fast fading, see e.g., \cite{Komninakis_7, Santipach_8, Taubock_9, Bjornson_10, Choi_11}. Specifically, the estimation accuracy in a temporally correlated channel can be improved by employing the Kalman filter \cite{Komninakis_7, Santipach_8}. Compressed sensing can be utilized to optimize the delay-Doppler basis of a doubly selective fading channel to improve the estimation accuracy \cite{Taubock_9}. However, these methods do not aim to reduce the estimation overhead, i.e., the amount of pilots used for channel estimation \cite{Komninakis_7, Santipach_8, Taubock_9}. On the other hand, for a spatially correlated channel, it has been indicated in \cite{Bjornson_10,Choi_11} that the pilot size can be reduced if the number of statistical dominant subspaces is smaller than the number of transmit antennas, at the cost of losing some multiplexing gain. Summarily, it remains a challenging problem to reduce the pilot overhead for large-scale MIMO systems in a high-speed environment.

On the other hand, due to the advances in indoor and outdoor positioning techniques, the real-time position information of the mobile terminal can be made available. In several prior applications, position information has been already used for routing \cite{Stojmenovic_12}, clustering \cite{Wang_13}, resource allocation \cite{Li_14,Xu_15}, etc. For high-speed railway communications, \cite{Liu_16} proposed a position-based channel model and \cite{Ghazal_17} extended the concept to multi-antenna wireless links. Further, position information was utilized to improve the channel estimation accuracy of high-speed railway communications in \cite{Ren_19}.
An interesting phenomenon caused by the mobility, called the joint spatial-temporal correlation, was discussed in \cite{Sternad_26,Jamaly_27,Zhang_28,Han_29}. It characterizes the relationship between the channel realizations of distinct antenna pairs at different time due to the mobility of antenna array. In particular, some measurement results between the BS and vehicles with multiple antennas were provided in \cite{Sternad_26}. \cite{Jamaly_27} discussed the effect of the mutual electromagnetic coupling between different antenna elements. \cite{Zhang_28} proposed a novel differential modulation for the moving antenna array based on it. \cite{Han_29} discussed the application of spatio-temporal correlation in reducing handover frequency in high-speed railway scenario.

In this paper, we focus on the training-based channel estimation in a large-scale MIMO system under high-speed railway scenarios. It is assumed that the BS is static and the train moves linearly with constant velocity, both employing linear antenna arrays. We mainly consider the uplink channel estimation, while the results can also be used for the downlink due to the channel reciprocity. We find that the joint spatial-temporal correlation can be utilized to significantly reduce the estimation overhead with the help of position information and then propose a position-aided channel estimator. It will be shown that its performance deteriorates a little as the mobility increases.
More specifically, during the training phase of each data block, it is better to select a subset of the transmit antennas to send pilot symbols and an initial estimate of the channel submatrix corresponding to this part of transmit antennas can be utilized repeatedly. Later, the estimate of the entire channel matrix could be constructed based on the initial submatrix and the location information of the transmit antenna array, by exploiting the spatial-temporal correlation of the channel. We then analyze its performance in term of the achievable throughput. Finally, we present the optimal selections of system parameters including power allocation, training interval and antenna size, by maximizing the obtained achievable throughput bound in this paper.

It is worth noting that the joint spatial-temporal correlation is significantly different from the conventional spatial correlation or temporal correlation \cite{Zhang_28}. In this paper, we assume that the antennas are sufficiently separated, so there is no spatial correlation between antenna elements. Besides, under the highly mobile condition, the coherent interval of the channel is so small that the temporal correlation is very weak. The spatial-temporal correlation here refers to the fact that due to the high mobility, the channel responses of different antenna pairs along the moving path at different time are correlated. Hence, the methods and the results based on conventional spatially correlated channel (such as \cite{Bjornson_10}) can not be applied directly here.

The remainder of this paper is organized as follows. The channel model is introduced in Section II, where the joint spatial-temporal correlation is presented. Then, the position-aided channel estimator is developed in Section III. In Section IV, the performance of the system with the new proposed training scheme is analyzed and the optimal system parameter selections are presented. Simulation results are given in Section V. Finally, conclusions are drawn in Section VI.


\begin{table}
\centering
\caption{Some important variables for problem description in this paper.}\label{table one}
\begin{tabular}{|c|c|c|}
\hline
Variable  & Description   \\
\hline
$M,N$ & The numbers of transmit antennas and receive antennas  \\
\hline
$H_k$ & The channel state matrix during the \mbox{$k$-th} signal block \\
\hline
$h_{n,m}(k)$ & The channel state between \mbox{$m$-th} transmit antenna and \mbox{$n$-th} receive antenna \\
\hline
$\bm{h}_k^m$ & The channel state vector between \mbox{$m$-th} transmit antenna and all receive antennas  \\
\hline
$z_k^m$ & The position of \mbox{$m$-th} transmit antenna during the \mbox{$k$-th} signal block \\
\hline
$\eta$ & The correlation coefficient between different channel state vector  \\
\hline
$\theta$ & The moving direction of the terminal with respect to the line-of-sight direction\\
\hline
$\psi$ & The direction of linear antenna array with respect to the line-of-sight direction \\
\hline
$J_0(\cdot)$ & The zero-th order Bessel function of the first kind \\
\hline
$T_0$ & The length of each signal block\\
\hline
$t_0$ & The coherence time of the environment\\
\hline
\end{tabular}
\end{table}

\section{Channel Model}
\begin{figure}[!t]
\centering
\includegraphics[width=3 in]{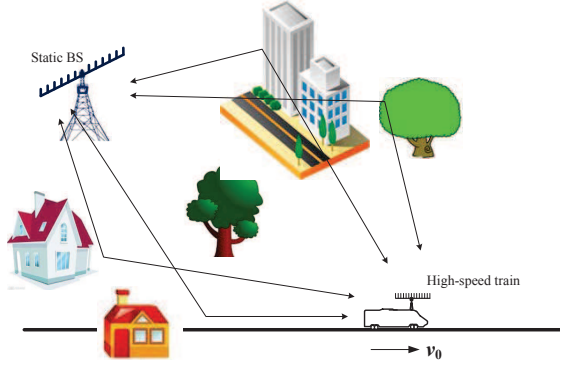}
\caption{The large-scale MIMO communication system for high-speed railways.}
\label{fig:system_structure_diagram}
\end{figure}

As shown in Fig. \ref{fig:system_structure_diagram}, we consider a point-to-point highly mobile large-scale MIMO system in a high-speed railway, where the BS is static and the terminal is in linear uniform motion with constant velocity $v_0$. Suppose that the channel is reciprocal, we concentrate on the uplink channel estimation problem and the results can be directly used in the downlink. According to the training-based system architecture, each signal block is divided into two parts: training phase and data phase. Some known training symbols are sent by the transmitter to estimate the CSI during the training phase and then the estimated channel is used in the following data phase. It is assumed that the channel state keeps constant during the same block, and changes to other values between different blocks. Besides, let the carrier wavelength be $\lambda_0$ and symbol rate be $B_0$, then the maximum Doppler shift is $f_D=\frac{v_0}{\lambda_0}$, the coherence time of the channel is $\frac{\lambda_0 B_0}{2 v_0}$, and the length of each signal block is set as $T_0=\lfloor\frac{\lambda_0 B_0}{2\xi_0 v_0}\rfloor$ symbols (where the constant $\xi_0$ should satisfy $\xi_0 \gg 1$).

We assume that a linear antenna array is employed at the mobile terminal (i.e., the train). The number of transmit antennas and receive antennas are denoted as $M$ and $N$, respectively. We focus on the effects of small-scale fast fading, which is modeled as Rayleigh distribution in this paper. Let $\bm{H}_k \in \mathbb{C}^{N \times M}$ be the channel matrix for the \mbox{$k$-th} signal block with its elements $h_{n,m}(k)$ denoting the channel state between the \mbox{$n$-th} receive antenna and the \mbox{$m$-th} transmit antenna (where $h_{n,m}(k)\sim \mathcal{CN}(0,1)$). It is assumed that the distance between adjacent antennas is $\frac{\lambda_0}{2}$, so there is no spatial correlation between the antenna elements and the elements in $\bm{H}_k$ are i.i.d. Further, let $\bm{h}_k^m \in \mathbb{C}^{N \times 1}$ denote the channel vector between the \mbox{$m$-th} transmit antenna and all $N$ receive antennas in the \mbox{$k$-th} block, namely $\bm{H}_k=[\bm{h}_k^1, \bm{h}_k^2,... , \bm{h}_k^{M}]$. Consequently, these $M$ channel vectors are independent of each other.

\begin{figure}[!t]
\centering
\includegraphics[width=5.5 in]{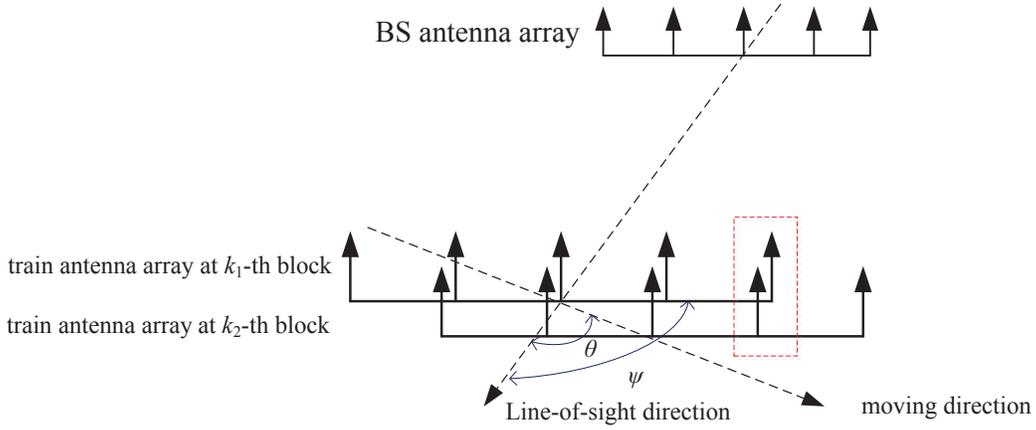}
\caption{The joint spatio-temporal correlation of moving antenna array \cite{Zhang_28}.}
\label{fig:spatio_temporal_correlation}
\end{figure}

Next we introduce the concept of joint spatio-temporal correlation. As shown in Fig. \ref{fig:spatio_temporal_correlation}, the moving direction of the terminal is $\theta$ with respect to the line-of-sight direction, and the direction of the linear antenna array is $\psi$. Fig. \ref{fig:spatio_temporal_correlation} depicts the specific locations of the entire moving antenna array of transmitter at the \mbox{$k_1$-th} and \mbox{$k_2$-th} signal blocks. It can be seen that the first antenna of the transmitter to the right at the \mbox{$k_1$-th} block is located at nearly the same place as the second transmit antenna at the \mbox{$k_2$-th} block due to the mobility of terminal. The corresponding channel vectors are $\bm{h}_{k_1}^1$ and $\bm{h}_{k_2}^2$. Intuitively, there exists some correlation between $\bm{h}_{k_1}^1$ and $\bm{h}_{k_2}^2$ according to many channel models, such as the Clark's model \cite[Sec 2.4]{Jakes_21}. Such correlation is termed as joint spatio-temporal correlation, which captures the correlation between distinct antenna pairs at different time due to mobility.
In general, the specific correlation between $\bm{h}_{k_1}^1$ and $\bm{h}_{k_2}^2$ can be estimated from measurement. Here, we introduce an analytical model. Specifically, when the moving scattering objects are modeled by poisson point process, the final correlation coefficient between $\bm{h}_{k_1}^1$ and $\bm{h}_{k_2}^2$ can be expressed as follows (see more details in \cite{Zhang_28,Abdi_33} and the measurements can be found in \cite{Sternad_26,Jamaly_27})
\begin{equation}\label{Eq:the final correlation coefficient}
  \eta=\frac{J_0(\sqrt{a^2+b^2-\kappa^2-2ab \cos(\psi-\theta)+j2 \kappa[a\cos(\mu-\theta)+b\cos(\mu-\psi)]})}{J_0(\kappa)},
\end{equation}
where $J_0(\cdot)$ is the zero-th order Bessel function of the first kind; $\kappa$ indicates the width of angle of the arrival (AOA) and $\mu$ accounts the mean direction of AOA; $a=2 \pi f_D \tau$ and $b=2 \pi D/ \lambda_0$, with $\tau$ being the time interval between the \mbox{$k_1$-th} and \mbox{$k_2$-th} blocks, and $D$ being the antenna spacing.

Assuming that the mobile terminal can adapt the direction of antenna array so that $\psi=\theta$ to achieve the largest correlation. And the scattering is isotropic so we have $\kappa=0$. Hence, the correlation can be simplified as
\begin{equation}\label{Eq:the simplified correlation coefficient}
  \eta=J_0(\sqrt{a^2+b^2-2ab \cos(\psi-\theta)})=J_0\Big(\frac{2\pi |v_0 \tau -D|}{\lambda_0}\Big).
\end{equation}

Denote the location of the first transmit antenna at the \mbox{$k_1$-th} block as $z_{k_1}^1$ and the location of second antenna at the \mbox{$k_2$-th} block as $z_{k_2}^2$. Then, we have $|v_0 \tau -D|=|z_{k_2}^2-z_{k_1}^1|$.
Consequently, we can extend (\ref{Eq:the simplified correlation coefficient}) as the correlation expression between the response of first antenna at the \mbox{$k_1$-th} block $\bm{h}_{k_1}^1$ and the response of \mbox{$m$-th} antenna at the \mbox{$k_m$-th} block $\bm{h}_{k_m}^m$ as
\begin{equation}\label{eqn:temporal correlation coefficient}
  \eta(z_{k_1}^1,z_{k_m}^m)= J_0 \big( \frac{2\pi|z_{k_m}^m-z_{k_1}^1|}{\lambda_0} \big).
\end{equation}

We assume that the relative position of the transmit antenna array is precisely known at any time, so is the correlation in (\ref{eqn:temporal correlation coefficient}). Besides, we make the following assumption.

\emph{\textbf{Assumption 1:} The channel state at a fixed position within the fading field stays constant during a period $t_0$ and after that may change to some other value, where $t_0$ is called the coherence time of the environment and determined by the time variation of the scatterers.}

\emph{\textbf{Remark 1:} }It is worth noting that the channel coherence time $\frac{\lambda_0}{2v_0}$ and the environment coherence time $t_0$ are fundamentally different. The former is determined by the moving speed of the transmitter while the latter is by the time variation of the scatterers in the radio propagation paths. In general, $\frac{\lambda_0}{2v_0} \ll t_0$ since the environment can not change much within a short period.


\section{Position-aided Channel Estimation}

In the conventional approach, the entire channel matrix is re-estimated in each block, to cope with the channel variation caused by high mobility. Thereby, in order to estimate the channel vectors of $M$ transmit antennas, at least $M$ pilot symbols need to be transmitted during training phase, which leads to huge training overhead in a large-scale MIMO system \cite{Hassibi_5}.
To reduce the training overhead, we propose a new channel estimation concept, called position-aided channel estimation, by exploiting the property of joint spatio-temporal correlation. It is assumed that all transmit antennas form a linear array with uniform interval $\frac{\lambda_0}{2}$ and that they move along the same path. Then during the training phase of each block, we have to only estimate the channel vectors of a subset of the transmit antennas by transmitting pilot symbols, while the rest of the channel vectors can be obtained through linear interpolation based on the joint spatial-temporal correlation. As a result, the overhead of the training stage of each block can be significantly reduced, resulting in high data throughput.

\subsection{Initial Estimation of the First Column in Each Group Based on Pilots}

Let $\bm{h}_k^m \in \mathbb{C}^{N \times 1}$ denote the channel vector between the \mbox{$m$-th} transmit antenna and all $N$ receive antennas in the \mbox{$k$-th} block, thus $\bm{H}_k=[\bm{h}_k^1, \bm{h}_k^2,... , \bm{h}_k^{M}]$.
These $M$ channel vectors are further divided into $M_g$ groups, each containing $M_e$ adjacent columns in $\bm{H}_k$. Thus, $M=M_e \cdot M_g$. Then, the channel sub-matrix for the \mbox{$i$-th} group can be expressed as
\begin{equation}\label{eqn:the expression of each group}
  \bm{Q}_{k}^{i}=[\bm{q}_{k,i}^1,\bm{q}_{k,i}^2,...,\bm{q}_{k,i}^{M_e}]=[\bm{h}_{k}^{M_e(i-1)+1)}, \bm{h}_{k}^{M_e(i-1)+2)},..., \bm{h}_{k}^{M_e(i-1)+M_e)}],\,\,i=1,2,...,M_g.
\end{equation}
And the channel matrix $\bm{H}_k$ can be rewritten as
\begin{equation}\label{eqn:expression of channel matrix}
  \bm{H}_k=[\bm{Q}_{k}^{1},\bm{Q}_{k}^{2},...,\bm{Q}_{k}^{M_g}].
\end{equation}

Under the position-aided channel estimation scheme, only the first transmit antenna in each group sends pilot symbols to estimate the channel state during each block, which corresponds to the following $N \times M_g$ sub-matrix of $\bm{H}_k$:
\begin{equation}\label{eqn:subset of channel matrix}
  \bm{G}_k=[\bm{g}_{k}^1,\bm{g}_{k}^2,...,\bm{g}_{k}^{M_g}]=[\bm{h}_{k}^{1}, \bm{h}_k^{M_e+1},... , \bm{h}_k^{M_e(M_g-1)+1}].
\end{equation}

Denote $T_\tau$ as the training duration in terms of the number of pilot symbols, and let $\bm{S}_{\tau,k}\in \mathbb{C}^{M_g \times T_\tau}$ and $\bm{Y}_{\tau,k}\in \mathbb{C}^{N \times T_\tau}$ be the pilot symbol matrix and the corresponding received signal during the training phase, respectively. Then, the training phase can be modeled as
\begin{equation}\label{eqn:training phase}
  \bm{Y}_{\tau,k}=\sqrt{\frac{P_\tau}{M_g}} \bm{G}_k \bm{S}_{\tau,k}+\bm{V}_{\tau,k},
\end{equation}
where $P_\tau$ is the transmit power during the training phase and $\bm{V}_{\tau,k}\in \mathbb{C}^{N \times T_\tau}$ represents additive white Gaussian noise with i.i.d. $\mathcal{CN}(0,1)$ elements.

The minimum mean-square error (MMSE) estimate of $\bm{{G}}_k$ is given by
\begin{equation}\label{eqn:estimation result under MMSE in new system}
  \bm{\widehat{G}}_k=\sqrt{\tfrac{M_g}{P_\tau}} \bm{Y}_{\tau,k} \bm{S}_{\tau,k}^H \Big(\tfrac{M_g}{P_\tau} \bm{I}_{M_g}+\bm{S}_{\tau,k} \bm{S}_{\tau,k}^H \Big)^{-1}.
\end{equation}
With orthogonal pilot symbol sequences, i.e., $\bm{S}_{\tau,k} \bm{S}_{\tau,k}^H=\bm{I}_{M_g}T_\tau$, substituting (\ref{eqn:training phase}) into (\ref{eqn:estimation result under MMSE in new system}), we get
\begin{equation}\label{eqn:rewritten estimation result under MMSE in new system}
  \bm{\widehat{G}}_k=\frac{\frac{P_\tau T_\tau}{M_g}}{1+\frac{P_\tau T_\tau}{M_g}} \bm{G_k}+ \frac{\sqrt{\frac{P_\tau T_\tau}{M_g}}}{1+\frac{P_\tau T_\tau}{M_g}} \bm{V'}_{\tau,k},
\end{equation}
where $\bm{V'}_{\tau,k}=\frac{1}{\sqrt{T_\tau}}\bm{V}_{\tau,k}\bm{S}_{\tau,k}^H$, the elements of which are still i.i.d. $\mathcal{CN}(0,1)$.

Let $\bm{\widehat{G}}_k= [\bm{\widehat{g}}_{k}^1, \bm{\widehat{g}}_{k}^2,..., \bm{\widehat{g}}_{k}^{M_g}]$ and $\bm{V'}_{\tau,k}= [\bm{v}_{k}^1, \bm{v}_{k}^2,..., \bm{v}_{k}^{M_g}]$. Then
\begin{equation}\label{eqn:estimation result over moving path for i-th group}
  \widehat{\bm{g}}_k^i
  =\frac{\frac{P_\tau T_\tau}{M_g}}{1+\frac{P_\tau T_\tau}{M_g}} \bm{g}_k^i+ \frac{\sqrt{\frac{P_\tau T_\tau}{M_g}}}{1+\frac{P_\tau T_\tau}{M_g}} \bm{v}_{k}^i, \,\,i=1,2,...,M_g.
\end{equation}
Hence, we have these initial estimates of the first channel column vectors that are independent and identical distributed as
\begin{equation}\label{eqn:estimation error for the first column}
  \bm{\widehat{g}}_k^i \sim \mathcal{CN} \Big( 0, \frac{\frac{P_\tau}{M_g}T_\tau}{1+\frac{P_\tau}{M_g}T_\tau} \bm{I}_{N}\Big), \,\,i=1,2,...,M_g.
\end{equation}

\begin{figure}[!t]
\centering
\includegraphics[width=4.5 in]{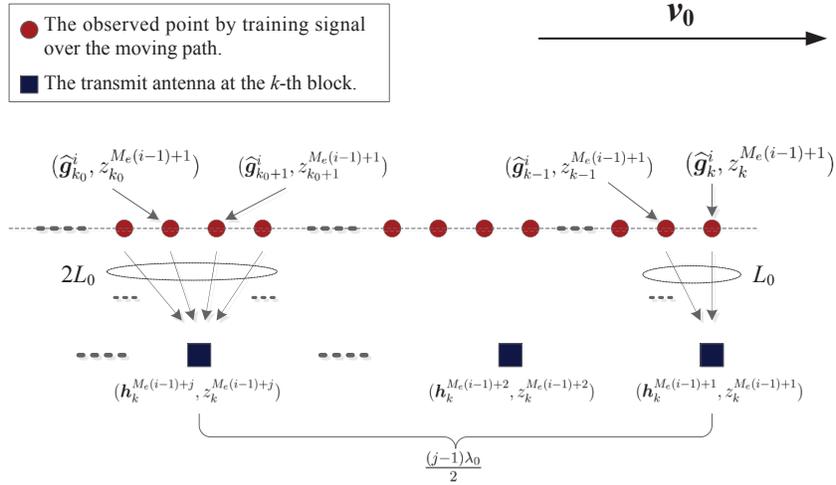}
\caption{The diagram of the estimation process for the \mbox{$i$-th} group under the proposed position-aided channel estimation scheme.}
\label{fig:basic_idea_of_this_paper}
\end{figure}

Let $z_k^{(M_e(i-1)+1)}$ be the position of the first transmit antenna in the \mbox{$i$-th} group of the \mbox{$k$-th} signal block over the moving path. Then, $\widehat{\bm{g}}_k^i$ can be regarded as the CSI sample at the point $z_k^{(M_e(i-1)+1)}$ on the moving path.
As shown in Fig. \ref{fig:basic_idea_of_this_paper}, a group of CSI samples along the moving path can be obtained over time \mbox{$k$}. We then establish the following CSI table $\bm{\Phi}_{k}^i$ for the \mbox{$i$-th} group
\begin{equation}\label{eqn:CSI tables for i group}
  \bm{\Phi}_{k}^i=\big\{(\bm{\widehat{g}}_l^{i},z_l^{(M_e(i-1)+1)}), \,l=1,2,...,k \big\}, \,\,i=1,2,...,M_g,
\end{equation}
which can be used to obtain the estimates of all channel vectors in the \mbox{$i$-th} group, i.e., $\bm{\widehat{Q}}_k^i$. The details of this will be given in the sequel.

\subsection{Refined Estimation of the First Column in Each Group}

As shown in Fig. \ref{fig:basic_idea_of_this_paper}, we will use $L_0$ samples $\{\bm{\widehat{g}}_{k-L_0+1}^{i},..., \bm{\widehat{g}}_{k-1}^{i},\bm{\widehat{g}}_k^{i}\}$ over the moving path to refine the estimate of $\bm{{h}}_{k}^{(M_e(i-1)+1)}=\bm{{g}}_k^i$ with the help of position information (where $L_0 \leq \xi_0$ due to the constraint of channel coherence distance). For notational simplicity, we denote $i_0=M_e(i-1)+1$ in this subsection.

As stated in the previous section, $\bm{g}_{k-L_0+1}^{i},... ,\bm{g}_{k-1}^{i}$ and $\bm{g}_k^{i}$ are jointly Gaussian distributed with zero mean and the following covariance matrix: 
\begin{equation}\label{eqn:covariance matrix in first column estimation}
\begin{split}
\bm{R}_{\bm{g}_{k}^i,\bm{g}_{k-1}^i,...,\bm{g}_{k-L_0+1}^i}=&
\left(                 
  \begin{array}{cccc}   
    \bm{R}_{\bm{g}_{k}^i} & \bm{R}_{\bm{g}_{k}^i\bm{g}_{k-1}^i} & \cdots & \bm{R}_{\bm{g}_{k}^i\bm{g}_{k-L_0+1}^i}\\  
    \bm{R}_{\bm{g}_{k-1}^i\bm{g}_{k}^i} & \bm{R}_{\bm{g}_{k-1}^i} & \cdots & \bm{R}_{\bm{g}_{k-1}^i\bm{g}_{k-L_0+1}^i}\\  
    \vdots & \, & \ddots & \, \\  
    \bm{R}_{\bm{g}_{k-L_0+1}^i\bm{g}_{k}^i} & \bm{R}_{\bm{g}_{k-L_0+1}^i\bm{g}_{k-1}^i} & \cdots & \bm{R}_{\bm{g}_{k-L_0+1}^i}  
  \end{array}
\right) \\                
=&
\left(                 
  \begin{array}{cccc}   
    \eta_{1,1} & \eta_{1,2} & \cdots & \eta_{1,L_0}\\  
    \eta_{2,1} & \eta_{2,2} & \cdots & \eta_{2,L_0} \\  
    \vdots & \, & \ddots & \, \\  
    \eta_{L_0,1} & \eta_{L_0,2} & \cdots & \eta_{L_0,L_0}  
  \end{array}
\right)                 
\otimes
\bm{I}_N,
\end{split}
\end{equation}
where $\otimes$ denotes the Kronecker product and $\bm{R}_{\bm{x}\bm{y}}=\mathbb{E} \{(\bm{x}-\mathbb{E}[\bm{x}])(\bm{y}-\mathbb{E}[\bm{y}])^H\}$. By plugging the location expressions into (\ref{eqn:temporal correlation coefficient}), we have
\begin{equation}\label{eqn:the expression of correlation}
  \eta_{m,n}=J_0\big(2\pi |m-n| \frac{v_0 T_0}{\lambda_0 B_0}\big).
\end{equation}

However, the receiver does not know the exact values of $\{\bm{g}_{k-L_0+1}^{i},..., \bm{g}_{k-1}^{i}, \bm{g}_k^{i}\}$, but only has their initial estimates based on pilot symbols, namely $\{\bm{\widehat{g}}_{k-L_0+1}^{i},..., \bm{\widehat{g}}_{k-1}^{i},\bm{\widehat{g}}_k^{i}\}$. Based on (\ref{eqn:estimation result over moving path for i-th group}), $\{\bm{\widehat{g}}_{k-L_0+1}^{i}, \bm{\widehat{g}}_{k-L_0+2}^{i}, ..., \bm{\widehat{g}}_{k}^{i},\bm{g}_k^{i}\}$ are also jointly Gaussian distributed with covariance matrix (recall that $\bm{h}_{k}^{i_0}=\bm{g}_k^{i}$)
\begin{equation}\label{eqn:covariance matrix in first column estimation 2}
\begin{split}
\bm{R}_{\bm{h}_{k}^{i_0},\bm{\widehat{g}}_{k}^i,...,\bm{\widehat{g}}_{k-L_0+1}^i}=
& \left(                 
  \begin{array}{ccccc}   
    \bm{R}_{\bm{g}_{k}^i} & \bm{R}_{\bm{g}_{k}^i \bm{\widehat{g}}_{k}^i} & \bm{R}_{\bm{g}_{k}^i\bm{\widehat{g}}_{k-1}^i} & \cdots & \bm{R}_{\bm{g}_{k}^i\bm{\widehat{g}}_{k-L_0+1}^i}\\  
    \bm{R}_{\bm{\widehat{g}}_{k}^i\bm{g}_{k}^i} & \bm{R}_{\bm{\widehat{g}}_{k}^i} & \bm{R}_{\bm{\widehat{g}}_{k}^i \bm{\widehat{g}}_{k-1}^i} & \cdots & \bm{R}_{\bm{\widehat{g}}_{k}^i \bm{\widehat{g}}_{k-L_0+1}^i}\\  
    \bm{R}_{\bm{\widehat{g}}_{k-1}^i\bm{g}_{k}^i} & \bm{R}_{\bm{\widehat{g}}_{k-1}^i\bm{\widehat{g}}_{k}^i} & \bm{R}_{\bm{\widehat{g}}_{k-1}^i}  & \cdots & \bm{R}_{\bm{\widehat{g}}_{k-1}^i\bm{\widehat{g}}_{k-L_0+1}^i}\\  
    \vdots & \, & \, &  \ddots & \, \\  
    \bm{R}_{\bm{\widehat{g}}_{k-L_0+1}^i\bm{g}_{k}^i} & \bm{R}_{\bm{\widehat{g}}_{k-L_0+1}^i\bm{\widehat{g}}_{k}^i} & \bm{R}_{\bm{\widehat{g}}_{k-L_0+1}^i\bm{\widehat{g}}_{k-1}^i} & \cdots & \bm{R}_{\bm{\widehat{g}}_{k-L_0+1}^i}  
  \end{array}
\right)             \\    
= &
\left(                 
  \begin{array}{ccccc}   
    1 & \eta_{1,1}\sigma_0^2 & \eta_{1,2}\sigma_0^2 & \cdots & \eta_{1,L_0}\sigma_0^2\\  
    \eta_{1,1}\sigma_0^2 & \eta_{1,1}\sigma_0^2 & \eta_{1,2}\sigma_0^4 & \cdots & \eta_{1,L_0}\sigma_0^4 \\  
    \eta_{2,1}\sigma_0^2 & \eta_{2,1}\sigma_0^4  & \eta_{2,2}\sigma_0^2 & \cdots & \eta_{2,L_0}\sigma_0^4\\  
    \vdots & \, & \, & \ddots & \, \\  
    \eta_{L_0,1}\sigma_0^2 & \eta_{L_0,1}\sigma_0^4 & \eta_{L_0,2}\sigma_0^4 & \cdots & \eta_{L_0,L_0}\sigma_0^2  
  \end{array}
\right)                 
\otimes
\bm{I}_N,
\end{split}
\end{equation}
where
\begin{equation}\label{eqn:the expression of sigma}
  \sigma_0^2=\frac{\frac{P_\tau T_\tau}{M_g}}{1+\frac{P_\tau T_\tau}{M_g}}.
\end{equation}

Given the initial estimates in the CSI table $\bm{\Phi}_{k}^i$, the MMSE estimate of $\bm{h}_{k}^{i_0}$ is \cite{Kay_24}
\begin{equation}\label{eqn:MMSE estimator for the first column in each group}
\begin{split}
  \bm{\widehat{h}}_{k}^{i_0}= &\mathbb{E}\big\{\bm{h}_{k}^{i_0}\big| \bm{\widehat{g}}_{k-L_0+1}^{i},\bm{\widehat{g}}_{k-L_0+2}^{i},..., \bm{\widehat{g}}_{k}^{i}\big\}  \\
  = &
\left(                 
  \begin{array}{c}   
    \eta_{1,1}\sigma_0^2\\  
    \eta_{1,2}\sigma_0^2\\  
    \vdots \\  
    \eta_{1,L_0}\sigma_0^2\\  
  \end{array}
\right)^T
\left(                 
  \begin{array}{cccc}   
    \eta_{1,1}\sigma_0^2 & \eta_{1,2}\sigma_0^4 & \cdots & \eta_{1,L_0}\sigma_0^4 \\  
    \eta_{2,1}\sigma_0^4  & \eta_{2,2}\sigma_0^2 & \cdots & \eta_{2,L_0}\sigma_0^4\\  
    \vdots & \, & \ddots & \, \\  
    \eta_{L_0,1}\sigma_0^4 & \eta_{L_0,2}\sigma_0^4 & \cdots & \eta_{L_0,L_0}\sigma_0^2  
  \end{array}
\right)^{-1}
\otimes \bm{I}_N
\left(                 
  \begin{array}{c}   
    \bm{\widehat{g}}_{k}^i\\  
    \bm{\widehat{g}}_{k-1}^i\\  
    \vdots \\  
    \bm{\widehat{g}}_{k-L_0+1}^i\\  
  \end{array}
\right).
\end{split}
\end{equation}
And the corresponding MMSE matrix is \cite{Kay_24}
\begin{equation}\label{eqn:mean square estimation error for the first column}
\begin{split}
  & \mathbb{E}\Big\{\big(\bm{h}_{k}^{i_0}-\bm{\widehat{h}}_{k}^{i_0}\big) \big(\bm{h}_{k}^{i_0}-\bm{\widehat{h}}_{k}^{i_0}\big)^H\Big\} \\
  = &\bm{I}_N-
\left(                 
  \begin{array}{c}   
    \eta_{1,1}\sigma_0^2\\  
    \eta_{1,2}\sigma_0^2\\  
    \vdots \\  
    \eta_{1,L_0}\sigma_0^2\\  
  \end{array}
\right)^T
\left(                 
  \begin{array}{cccc}   
    \eta_{1,1}\sigma_0^2 & \eta_{1,2}\sigma_0^4 & \cdots & \eta_{1,L_0}\sigma_0^4 \\  
    \eta_{2,1}\sigma_0^4  & \eta_{2,2}\sigma_0^2 & \cdots & \eta_{2,L_0}\sigma_0^4\\  
    \vdots & \, & \ddots & \, \\  
    \eta_{L_0,1}\sigma_0^4 & \eta_{L_0,2}\sigma_0^4 & \cdots & \eta_{L_0,L_0}\sigma_0^2  
  \end{array}
\right)^{-1}
\left(                 
  \begin{array}{c}   
    \eta_{1,1}\sigma_0^2\\  
    \eta_{2,1}\sigma_0^2\\  
    \vdots \\  
    \eta_{L_0,1}\sigma_0^2\\  
  \end{array}
\right)\otimes \bm{I}_N.
\end{split}
\end{equation}

Specifically, when $L_0=1$, the MMSE estimate in (\ref{eqn:MMSE estimator for the first column in each group}) can be simplified to
\begin{equation}\label{eqn:estimation result for the first column of i-th group}
  \bm{\widehat{h}}_{k}^{i_0}
  =\bm{\widehat{g}}_{k}^{i}, \,\,i=1,2,...,M_g,
\end{equation}
with the MMSE matrix given in (\ref{eqn:mean square estimation error for the first column}) simplified to
\begin{equation}\label{eqn:mean square estimation error for first column}
  \mathbb{E} \Big\{\big( \bm{h}_{k}^{i_0}- \bm{\widehat{h}}_{k}^{i_0}\big) \big(\bm{h}_{k}^{i_0}- \bm{\widehat{h}}_{k}^{i_0}\big)^H\Big\}
  = \big(1-\sigma_0^2\big) \bm{I}_N,\,\,i=1,2,...,M_g.
\end{equation}

\subsection{Estimation of Other Columns in Each Group}

For the \mbox{$j$-th} column ($j=2,3,...,M_e$) in the \mbox{$i$-th} group, the channel vector $\bm{h}_{k}^{(M_e(i-1)+j)}$ can also be estimated by using the CSI table $\bm{\Phi}_k^i$ and the antenna position information $z_{k}^{(M_e(i-1)+j)}$. For clarity, we denote $i_0={M}_e(i-1)+1$ and $j_0={M}_e(i-1)+j$ in this subsection.
As observed from Fig. \ref{fig:basic_idea_of_this_paper}, $z_{k}^{j_0}$ is located between $z_{k_0}^{i_0}$ and $z_{k_0+1}^{i_0}$ along the moving path, namely $|z_{k}^{i_0} - z_{k_0}^{i_0}| \geq |z_{k}^{i_0} -z_{k}^{j_0}| \geq |z_{k}^{i_0} -z_{k_0+1}^{i_0}|$. The value of $k_0$ can be obtained by comparing $z_{k}^{j_0}$ with $\{z_{l}^{i_0}, l=1,2,...,k\}$ that is contained in $\bm{\Phi}_k^i$. In particular, on the condition that the transmitter is in uniform motion with speed $v_0$ and the interval between two adjacent antennas is $\frac{\lambda_0}{2}$, $|z_{m}^{i_0}-z_{n}^{i_0}|= \frac{|m-n|v_0T_0}{B_0}$. Then, by using $|z_k^{j_0}-z_k^{i_0}|=\frac{(j-1)\lambda_0}{2}$, the above condition becomes
\begin{align}\label{eqn:the result for k-0}
  & \frac{(k-k_0)v_0 T_0}{B_0} \geq \frac{(j-1)\lambda_0}{2} \geq  \frac{(k-k_0-1)v_0 T_0}{B_0} \\
  & \Rightarrow  \,\,\,\, k_0=\Big\lfloor k-\frac{(j-1)\lambda_0B_0}{2v_0 T_0} \Big\rfloor.
\end{align}
It can be seen that the particular value of $k_0$ depends on the value of $\xi_0$ in the expression of $T_0=\frac{\lambda_0 B_0}{2\xi_0v_0}$. For instance, $k-k_0 \approx (j-1)\xi_0$, if the effect of the floor operator in (\ref{eqn:the result for k-0}) is ignored.

As shown in Fig. \ref{fig:basic_idea_of_this_paper}, the next task is to estimate $\bm{h}_{k}^{j_0}$ based on $\{\bm{\widehat{g}}_{k-L_0+1}^{i},..., \bm{\widehat{g}}_{k_0}^{i}, \bm{\widehat{g}}_{k_0+1}^{i},..., \bm{\widehat{g}}_{k_0+L_0}^{i}\}$ aided by the position information. Similarly as before, $\bm{\widehat{g}}_{k-L_0+1}^{i},..., \bm{\widehat{g}}_{k_0}^{i},\bm{\widehat{g}}_{k_0+1}^{i},...,\bm{\widehat{g}}_{k_0+L_0}^{i}$ and $\bm{h}_{k}^{j_0}$ are jointly Gaussian distributed, with the covariance matrix
\begin{equation}\label{eqn:covariance matrix in other column estimation}
\begin{split}
\bm{R}_{\bm{\widehat{g}}_{k_0+L_0}^{i},...,\bm{\widehat{g}}_{k_0+1}^{i},\bm{h}_{k}^{j_0}, \bm{\widehat{g}}_{k_0}^{i},...,\bm{\widehat{g}}_{k-L_0+1}^{i}}= &
\left(                 
  \begin{array}{ccccc}   
    \bm{R}_{\bm{\widehat{g}}_{k+L_0}^i} & \cdots & \bm{R}_{\bm{\widehat{g}}_{k+L_0}^i\bm{{h}}_{k}^{j_0}} & \cdots & \bm{R}_{\bm{\widehat{g}}_{k+L_0}^i\bm{\widehat{g}}_{k-L_0+1}^i}\\  
     \vdots & \ddots &  &  & \\  
    \bm{R}_{\bm{{h}}_{k}^{j_0}\bm{\widehat{g}}_{k+L_0}^i} & \cdots & \bm{R}_{\bm{{h}}_{k}^{j_0}}  & \cdots & \bm{R}_{\bm{{h}}_{k}^{j_0}\bm{\widehat{g}}_{k-L_0+1}^i}\\  
    \vdots & \, & \, &  \ddots & \, \\  
    \bm{R}_{\bm{\widehat{g}}_{k-L_0+1}^i\bm{\widehat{g}}_{k+L_0}^i} & \cdots & \bm{R}_{\bm{\widehat{g}}_{k-L_0+1}^i \bm{{h}}_{k}^{j_0}} & \cdots & \bm{R}_{\bm{\widehat{g}}_{k-L_0+1}^i}\\  
  \end{array}
\right)  \\
=&
\left(                 
  \begin{array}{ccc}   
    \bm{R}_1 & \bm{r}_1^H & \bm{R}_2\\  
    \bm{r}_1 & 1 & \bm{r}_2 \\  
    \bm{R}_2^H & \bm{r}_2^H & \bm{R}_1\\  
  \end{array}
\right)                 
\otimes
\bm{I}_N,               
\end{split}
\end{equation}
with
\begin{align}\label{eqn:the expression for R-1, R-2, r-1 and r-2}
  \bm{R}_1 &=
  \left(                 
  \begin{array}{cccc}   
    \eta_{1,1}\sigma_0^2 & \eta_{1,2}\sigma_0^4 & \cdots & \eta_{1,L_0}\sigma_0^4\\  
    \eta_{2,1}\sigma_0^4 & \eta_{2,2}\sigma_0^2 & \cdots & \eta_{2,L_0}\sigma_0^4 \\  
    \vdots & \, & \ddots & \, \\  
    \eta_{L_0,1}\sigma_0^4 & \eta_{L_0,2}\sigma_0^4 & \cdots & \eta_{L_0,L_0}\sigma_0^2  
  \end{array}
\right),         \\        
  \bm{R}_2 &=
  \left(                 
  \begin{array}{cccc}   
    \eta_{1,L_0+1}\sigma_0^4 & \eta_{1,L_0+2}\sigma_0^4 & \cdots & \eta_{1,2L_0}\sigma_0^4\\  
    \eta_{2,L_0+1}\sigma_0^4 & \eta_{2,L_0+2}\sigma_0^4 & \cdots & \eta_{2,2L_0}\sigma_0^4 \\  
    \vdots & \, & \ddots & \, \\  
    \eta_{L_0,L_0+1}\sigma_0^4 & \eta_{L_0,L_0+2}\sigma_0^4 & \cdots & \eta_{L_0,2L_0}\sigma_0^4  
  \end{array}
\right),           \\      
  \bm{r}_1 &= [\eta_{L_0}' \sigma_0^2,\eta_{L_0-1}' \sigma_0^2,...,\eta_{2}' \sigma_0^2,\eta_{1}' \sigma_0^2], \\
  \bm{r}_2 &= [\eta_{1}'' \sigma_0^2,\eta_{2}'' \sigma_0^2,...,\eta_{L_0-1}'' \sigma_0^2,\eta_{L_0}'' \sigma_0^2],
\end{align}
where
\begin{align}\label{eqn:the expression for eta_i' and i''}
  & \eta_l'=J_0\Big(2\pi \frac{|z_k^{j_0}-z_{k_0+l}^{i_0}|}{\lambda_0} \Big), \\
  & \eta_l''=J_0\Big(2\pi \frac{|z_k^{j_0}-z_{k_0-l+1}^{i_0}|}{\lambda_0} \Big).
\end{align}


The MMSE estimate of $\bm{h}_k^{j_0}$ is given by
\begin{equation}\label{eqn:conditional mean estimator for j column}
\begin{split}
  \bm{\widehat{h}}_{k}^{j_0}= & \mathbb{E}\big\{\bm{h}_{k}^{j_0}\big|\bm{\widehat{g}}_{k-L_0+1}^{i},..., \bm{\widehat{g}}_{k_0}^{i},\bm{\widehat{g}}_{k_0+1}^{i},...,\bm{\widehat{g}}_{k_0+L_0}^{i}\big\}  \\
  = &
\left(                 
  \begin{array}{cc}   
    \bm{r}_1 & \bm{r}_{2} \\  
  \end{array}
\right)
\left(                 
  \begin{array}{cc}   
    \bm{R}_1 & \bm{R}_2 \\  
    \bm{R}_2^T & \bm{R}_1 \\  
  \end{array}
\right)^{-1} \otimes \bm{I}_N
\left(                 
  \begin{array}{c}   
    \bm{\widehat{g}}_{k_0+L_0}^i \\  
    \vdots \\
    \bm{\widehat{g}}_{k_0-L_0+1}^i\\  
  \end{array}
\right).
\end{split}
\end{equation}

Again the estimate is a linear combination of the $2L_0$ samples in $\bm{\Phi}_k^i$ and the interpolation coefficients can be precomputed offline. The corresponding MMSE matrix is given by
\begin{equation}\label{eqn:mean square estimation error for the other column}
   \mathbb{E}\big\{(\bm{h}_{k}^{j_0}-\bm{\widehat{h}}_{k}^{j_0}) (\bm{h}_{k}^{j_0}-\bm{\widehat{h}}_{k}^{j_0})^H\big\}
  = \bm{I}_N-
\left(                 
  \begin{array}{cc}   
    \bm{r}_1 & \bm{r}_{2} \\  
  \end{array}
\right)
\left(                 
  \begin{array}{cc}   
    \bm{R}_1 & \bm{R}_2 \\  
    \bm{R}_2^H & \bm{R}_1 \\  
  \end{array}
\right)^{-1}
\left(                 
  \begin{array}{c}   
    \bm{r}_1 \\  
    \bm{r}_2\\  
  \end{array}
\right)\otimes \bm{I}_N.
\end{equation}

Specifically, for the case of $L_0=1$, $\bm{\widehat{h}}_{k}^{j_0}$ can be simplified as
\begin{equation}\label{eqn:MMSE estimator for m+1}
\begin{split}
  \bm{\widehat{h}}_{k}^{j_0}= & \mathbb{E}\big\{\bm{{h}}_{k}^{j_0} \big| \bm{\widehat{g}}_{k_0+1}^i, \bm{\widehat{g}}_{k_0}^i\big\}  \\
  = &
\left(                 
  \begin{array}{cc}   
    \eta_1' \sigma_0^2 & \eta_1'' \sigma_0^2 \\  
  \end{array}
\right)
\left(                 
  \begin{array}{cc}   
    \sigma_0^2 & \eta_1 \sigma_0^4 \\  
    \eta_1 \sigma_0^4 & \sigma_0^2 \\  
  \end{array}
\right)^{-1}
\left(                 
  \begin{array}{c}   
    \bm{\widehat{g}}_{k_0+1}^i \\  
    \bm{\widehat{g}}_{k_0}^i \\  
  \end{array}
\right) \\
  = & \frac{\eta_1''-\eta_1\eta_1'\sigma_0^2}{1-\eta_1^2\sigma_0^4}  \bm{\widehat{g}}_{k_0+1}^i + \frac{\eta_1'-\eta_1\eta_1''\sigma_0^2}{1-\eta_1^2\sigma_0^4}  \bm{\widehat{g}}_{k_0}^i,
\end{split}
\end{equation}
where $\eta_1=J_0\big(2\pi \frac{|z_{k_0+1}^{i_0}-z_{k_0}^{i_0}|}{\lambda_0}\big)$, $\eta_1'=J_0\big(2\pi \frac{|z_k^{j_0}-z_{k_0+1}^{i_0}|}{\lambda_0}\big)$ and $\eta_1''=J_0\big(2\pi \frac{|z_k^{j_0}-z_{k_0}^{i_0}|}{\lambda_0}\big)$ based on (\ref{eqn:the expression for eta_i' and i''}). Finally, the MMSE matrix in (\ref{eqn:mean square estimation error for the other column}) can be simplified as
\begin{equation}\label{eqn:mean square estimation error for other column}
  \mathbb{E}\big\{ \bm{\widetilde{h}}_{k}^{j_0} (\bm{\widetilde{h}}_{k}^{j_0})^H\big\}= \big(1-\sigma_0^2 \cdot \Gamma(\eta_1,\eta_1',\eta_1'')\big)\,\bm{I}_N,
\end{equation}
where
\begin{equation}\label{eqn:expression for sigma-2}
  \Gamma(\eta_1,\eta_1',\eta_1'')= \frac{{\eta_1'}^2+{\eta_1''}^2-2\eta_1\eta_1'\eta_1''\sigma_0^2}
  {1-{\eta_1}^2\sigma_0^4}.
\end{equation}

\emph{\textbf{Remark 2:}} It is worth noting again that there is no spatial correlation between antenna elements due to sufficiently separation. Namely, $\bm{h}_k^1, \bm{h}_k^2,... , \bm{h}_k^{M}$ are independent of each other, so the results of \cite{Bjornson_10} based on the correlation structure among $\bm{h}_k^1, \bm{h}_k^2,... , \bm{h}_k^{M}$ can not be applied. However, based on \emph{Assumption 1}, we can establish the relationship between $\bm{{h}}_{k}^{j_0}$ and $\bm{\widehat{g}}_{k_0+1}^i$, $\bm{\widehat{g}}_{k_0}^i$ (the past estimates of the first transmit antenna in the same group) by utilizing the joint spatio-temporal correlation in (\ref{eqn:temporal correlation coefficient}) with the help of position information of the transmit antenna array. Thus, we can get the estimator in (\ref{eqn:MMSE estimator for m+1}) and reduce the training overhead.

\subsection{Summary and Comments}

In summary, for the \mbox{$k$-th} signal block, the channel estimation process consists of the following steps:

\begin{itemize}
\item
Step 1: The first transmit antenna of each group transmits pilot symbols. The receiver computes the initial estimate $\bm{\widehat{G}}_k$ based on (\ref{eqn:rewritten estimation result under MMSE in new system}) and (\ref{eqn:estimation result over moving path for i-th group}) using the received signals and updates the $M_g$ CSI tables $\bm{\Phi}_{k}^i=\big[\bm{\Phi}_{k-1}^i,(\bm{\widehat{g}}_k^{i}, z_k^{(M_e(i-1)+1)})\big]$, $i=1,2,..., M_g$.
\item
Step 2: The estimate of the first column $\bm{h}_{k}^{(M_e(i-1)+1)}$ in the \mbox{$i$-th} sub-matrix  $\bm{Q}_k^i$ is refined by using (\ref{eqn:MMSE estimator for the first column in each group}), $i=1,2,...,M_g$.
\item
Step 3: The estimate of the \mbox{$j$-th} column $\bm{h}_{k}^{(M_e(i-1)+j)}$ in the \mbox{$i$-th} sub-matrix $\bm{Q}_{k}^i$, $j=2,3,... ,M_e$, is computed by using (\ref{eqn:conditional mean estimator for j column}). This yields the estimate of the entire channel matrix \mbox{$\bm{\widehat{H}}_k=[\bm{\widehat{h}}_{k}^{1}, \bm{\widehat{h}}_{k}^{2},..., \bm{\widehat{h}}_{k}^{M}]$}.
\end{itemize}

\emph{\textbf{Remark 3:}} (Maximum value for $M_e$) Since the channel state over the moving path is estimated by the pilot symbols from the first transmit antenna in a group and the estimation results are finally reused by the last antenna in the same group, the time interval between which is $\Delta T=\frac{(M_e-1) \lambda_0}{2v_0}$. In order to guarantee that the channel state at a fixed point does not change during this period, $\Delta T$ should be less than the coherence time of the transmission environment $t_0$, i.e., $\frac{(M_e-1)\lambda_0}{2 v_0} < t_0$. Thus, the maximum allowable value for $M_e$ can be expressed as
\begin{equation}\label{eqn:maximum number of antenna}
  M_e=\big\lfloor\frac{2 v_0 t_0}{\lambda_0}+1\big\rfloor.
\end{equation}

\emph{\textbf{Remark 4:}} It is seen that the value of $M_e$ is bounded, especially when the speed $v_0$ is low. It means that the size of the antenna array is limited if we only employ one group, which will lead to a low throughput. So, in order to support a larger size of antenna array, multiple groups should be employed. The optimal value of $M_g$ will be considered in Section IV.D.

\section{Performance Analysis and Throughput Optimization}

\subsection{Effective SNR Analysis}

Denote $\bm{\widetilde{H}}_k=\bm{H}_k-\bm{\widehat{H}}_k$ as the channel estimation error. The data phase in the \mbox{$k$-th} block is
\begin{equation}\label{eqn:transmission process in new proposed group training scheme}
  \bm{Y}_{d,k}=\sqrt{\frac{P_d}{M}} \bm{\widehat{H}}_k  \bm{S}_{d,k} + \sqrt{\frac{P_d}{M}} \bm{\widetilde{H}}_k \bm{S}_{d,k} +\bm{V}_{d,k},
\end{equation}
where $P_d$ denotes the transmit power in the data phase, $\bm{V}_{d,k}$ is an additive white Gaussian noise term with i.i.d. $\mathcal{CN}(0,1)$ elements, while $\bm{S}_{d,k}\in \mathbb{C}^{M \times T_d}$ and $\bm{Y}_{d,k}\in \mathbb{C}^{N \times T_d}$ are the transmitted signal and received signal, respectively.

Since $\bm{\widehat{H}}_k$ is an MMSE estimate, the error $\bm{\widetilde{H}}_k$ is uncorrelated with $\bm{\widehat{H}}_k$ due to the orthogonality principle \cite{Kay_24}. Let $\bm{E}_{d,k}=\sqrt{\tfrac{P_d}{M}} \bm{\widetilde{H}}_k \bm{S}_{d,k} +\bm{V}_{d,k}$ be an equivalent additive noise term that combines the effects of channel noise and channel estimation error. It follows that $\bm{E}_{d,k}$ is also zero mean and uncorrelated with $\bm{\widehat{H}}_k \bm{S}_{d,k}$. It is known that for uncorrelated additive noise, the worst distribution in terms of capacity is Gaussian \cite{Hassibi_5,Medard_22,Lapidoth_23}. Thus, on the condition that the transmitted signal satisfies $\mathbb{E}\{\bm{S}_{d,k} \bm{S}_{d,k}^H\}=T_d \bm{I}_M$, a lower bound on the capacity during the data phase can be expressed as
\begin{equation}\label{eqn:lower bound capacity expression in prior work}
  C_{\text{worst}} = 
  \mathbb{E} \Big\{  \log_2 \det\Big(\bm{I}_N+\frac{P_d}{M} \bm{R}_{\bm{E}}^{-1} \widehat{\bm{H}}_k \widehat{\bm{H}}_k^H\Big) \Big\},
\end{equation}
where $\bm{R}_{\bm{E}}=\frac{1}{T_d}\mathbb{E}\{\bm{E}_{d,k} \bm{E}_{d,k}^H\}=\frac{P_d}{M}\mathbb{E}\{ \widetilde{\bm{H}}_k \widetilde{\bm{H}}_k^H \}+ \bm{I}_N$.

In what follows, for tractability of analysis, we focus on the special case of $L_0=1$ and the performance of the case with $L_0>1$ will be examined via simulations in the next section. Specifically, when $L_0=1$, the estimation errors are given by (\ref{eqn:mean square estimation error for first column}) and (\ref{eqn:mean square estimation error for other column}). We have the following result.

\emph{\textbf{Proposition 1:} $\Gamma(\eta_1,\eta_1',\eta_1'')>1$ ($\Gamma(\cdot)$ defined in (\ref{eqn:expression for sigma-2})), if the signal to noise ratio (SNR) during the training phase is less than a certain threshold value, namely $\frac{P_\tau T_\tau}{M_g}<\frac{\Omega}{1-\Omega}$, where}
\begin{equation}\label{eqn:constraint in Lemma 2}
  \Omega=\min \limits_{ \{z_{k}^{j_0},j_0=1,2,...M\} }
  \frac{2\eta_1\eta_1'\eta_1''-\sqrt{4\eta_1^2{\eta_1'}^2{\eta_2''}^2-4\eta_1^2({\eta_1'}^2+{\eta_1''}^2-1)}}{2\eta_1^2}.
\end{equation}
\begin{IEEEproof}
  See Appendix A.
\end{IEEEproof}

In particular, for the typical system parameter with $\xi_0=20$, i.e., $T_0=\lfloor \frac{\lambda_0 B_0}{40v_0}\rfloor$, it can be obtained that $\Omega=0.999997609$ and the SNR threshold value $\frac{\Omega}{1-\Omega}=56.2$dB.
Hence, if the SNR value is below $56.2$dB, a very mild condition that always holds in practice, we can obtain \mbox{$\Gamma(\eta_1,\eta_1',\eta_1'')>1$}. Then, the MMSE error of the first column in each group given by (\ref{eqn:mean square estimation error for first column}) is larger than that of other columns given by (\ref{eqn:mean square estimation error for other column}).

\emph{\textbf{Remark 5:}}
As shown in Fig. \ref{fig:basic_idea_of_this_paper}, $2L_0$ samples are utilized to estimate $\bm{{h}}_{k}^{(M_e(i-1)+j)}$ based on (\ref{eqn:MMSE estimator for m+1}) while only $L_0$ samples are utilized to estimate $\bm{{h}}_{k}^{(M_e(i-1)+1)}$ based on (\ref{eqn:estimation result for the first column of i-th group}) due to the causality constraint. Hence, unless the SNR during training phase is extremely large, the estimation error of the first column is typically larger than that of others in the same group.

Thus, in practice, the estimation error for the first column in each group is the largest compared with that of other columns. We can then obtain a further lower bound on the capacity by assuming that the covariance of the estimation error of any column is $(1-\sigma_0^2)\bm{I}_N$. That is, we can use the following system model to lower bound the capacity of the original system in (\ref{eqn:transmission process in new proposed group training scheme}):
\begin{equation}\label{eqn:new lower bound system}
  \bm{Y}_{d,k}=\sqrt{\frac{P_d}{M}} \bm{\widehat{H}}_k'  \bm{S}_{d,k} + \sqrt{\frac{P_d}{M}} \bm{\widetilde{H}}_k' \bm{S}_{d,k} +\bm{V}_{d,k},
\end{equation}
where $\bm{\widehat{H'}}_k$ contains i.i.d. $\mathcal{CN}(0,\sigma_0^2)$ elements while $\bm{\widetilde{H'}}_k$ contains i.i.d. $\mathcal{CN}(0,1-\sigma_0^2)$ elements, and they are uncorrelated with each other.

For the model in (\ref{eqn:new lower bound system}), we have $\mathbb{E}\{ \widetilde{\bm{H'}}_k (\widetilde{\bm{H'}}_k)^H \}=M(1-\sigma_0^2) \bm{I}_N$ and
$\bm{R}_{\bm{E'}}=\frac{1}{T_d} \mathbb{E}\{\bm{E'}_{d,k} (\bm{E'}_{d,k})^{H} \} =  \frac{P_d}{M}\mathbb{E}\{ \widetilde{\bm{H'}}_k (\widetilde{\bm{H'}}_k)^{H} \}+ \bm{I}_N=[P_d(1-\sigma_0^2)+1] \bm{I}_N$.
Then, using (\ref{eqn:lower bound capacity expression in prior work}), the lower bound on the capacity of the system in (\ref{eqn:transmission process in new proposed group training scheme}) during data phase can be expressed as
\begin{equation}\label{eqn:lower bound capacity expression for new system}
  C_{\text{L}} =  \mathbb{E} \Big\{  \log_2 \det\Big(\bm{I}_N+\frac{P_d \sigma_0^2}{1+P_d(1-\sigma_0^2)} \cdot \frac{\overline{\bm{H}}_k \overline{\bm{H}}_k^H}{M}\Big) \Big\},
\end{equation}
where the normalized channel estimate is $\overline{\bm{H}}_k=\frac{\widehat{\bm{H}}_k'}{\sqrt{\sigma_0^2}}$, consisting of $\mathcal{CN}(0,1)$ elements. 

\subsection{End-to-End Throughput Optimization and System Parameter selections}

Taking the training stage into account, we can maximize the system throughput by optimally allocating the channel resources between the training and data phases. That is
\begin{equation}\label{eqn:transmission performance of new system}
  R_{\text{\tiny{L}}}=\max \limits_{P_d, T_d}\,  \mathbb{E} \Big\{\frac{T_0-T_\tau}{T_0} \cdot \log_2 \det\big(\bm{I}_N+\frac{P_d \sigma_0^2}{1+P_d(1-\sigma_0^2)} \cdot \frac{ \overline{\bm{H}}_k\overline{\bm{H}}_k^H}{M} \big)\Big\},
\end{equation}
where the pre-log factor $\frac{T_0-T_\tau}{T_0}$ accounts for the estimation cost of channel uses, while $P_d$ and $T_d$ satisfy the following constraints of total time slot and total transmission energy per block:
\begin{equation}\label{eqn:the constraint for channel resource allocation}
  T_0=T_\tau+T_d,\,\,\,P_0T_0=P_\tau T_\tau+P_dT_d.
\end{equation}

Substituting the expression of $\sigma_0^2$ into (\ref{eqn:transmission performance of new system}), the effective signal-to-noise ratio (SNR) can be expressed as
\begin{equation}\label{eqn:expression of effective snr}
  \rho_{\text{eff}}=\frac{P_d \sigma_0^2}{1+P_d(1-\sigma_0^2)}=
  \frac{P_d \frac{P_\tau}{M_g} T_\tau}{1+P_d+\frac{P_\tau}{M_g} T_\tau}.
\end{equation}

In order to maximize the right-hand side of (\ref{eqn:transmission performance of new system}) with respect to power allocation and the time interval partition, namely $\{P_\tau,P_d\}$ and $\{T_\tau,T_d\}$, we have the following two Lemmas.

\emph{\textbf{Lemma 1:} (Optimal Power Ratio) The optimal power ratio is given by $\frac{P_d}{P_0}=\alpha \frac{T_0}{T_d}$, where}
\begin{equation}\label{eqn:power allocation coefficient in the new system}
  \alpha=\frac{\sqrt{T_d(M_g+P_0T_0)}}{\sqrt{M_g(T_d+P_0T_0)}+\sqrt{T_d(M_g+P_0T_0})}.
\end{equation}
\begin{IEEEproof}
  See Appendix B.
\end{IEEEproof}

\emph{\textbf{Lemma 2:} (Optimal Time Interval Partition) The optimal length of the training interval under the optimal power allocation ratio is $M_g$ for all possible $P_0$ and $T_0$.}
\begin{IEEEproof}
  See Appendix C.
\end{IEEEproof}

Then, by substituting the results in \emph{Lemma 1} and \emph{Lemma 2} into (\ref{eqn:transmission performance of new system}), we obtain the following conclusion.

\emph{\textbf{Proposition 2:} In a training-based system with position-aided channel estimation, the lower bound on the throughput under the optimal channel resource allocation can be expressed as}
\begin{equation}\label{eqn:optimal capacity under proposed system in the new system}
  R_{\text{\tiny{L}}}\,=\,  \mathbb{E} \Big\{ \frac{T_0-M_g}{T_0} \cdot  \log_2 \,\det \big(\bm{I}_N+\rho_{\text{eff}}^{*} \frac{ \overline{\bm{H}}_k\overline{\bm{H}}_k^H}{M} \big) \Big\},
\end{equation}
\emph{where}
\begin{equation}\label{eqn:effective snr value in new system}
\begin{split}
  &\rho_{\text{eff}}^{*}=\frac{P_0^2T_0^2}{[\sqrt{M_g(T_0-M_g+P_0T_0)}+\sqrt{(T_0-M_g)(M_g+P_0T_0)}]^2},\\
  &T_0=\Big\lfloor\frac{\lambda_0 B_0}{2\xi_0 v_0}\Big\rfloor.
\end{split}
\end{equation}

\subsection{Special Case: $M_e=1$---Conventional Training Scheme}

If we set $M_e=1$ (and $M_g=M$), the position-aided channel estimator will reduce to the conventional channel estimator.
To serve as a baseline, let us analyze the performance of conventional training. Substituting $M_e=1$ and $M_g=M$ into (\ref{eqn:optimal capacity under proposed system in the new system}) and (\ref{eqn:effective snr value in new system}), we can obtain the corresponding performance of the conventional training scheme, which is consistent with the prior work \cite{Hassibi_5}.

\emph{\textbf{Corollary 1:} In a training-based system with conventional training, the lower bound on the throughput under well-designed system parameters can be expressed as}
\begin{equation}\label{eqn:optimal capacity in conventional system}
  R_{\text{\tiny{L}}}\,=\,  \mathbb{E} \Big\{ \frac{T_0-M}{T_0} \cdot  \log_2 \,\det \big(\bm{I}_N+\rho_{\text{eff}}^{*} \frac{ \overline{\bm{H}}_k\overline{\bm{H}}_k^H}{M} \big) \Big\},
\end{equation}
\emph{where}
\begin{equation}\label{eqn:effective snr value in conventional system}
  \rho_{\text{eff}}^{*}=\frac{P_0^2T_0^2}{[\sqrt{M(T_0-M+P_0T_0)}+\sqrt{(T_0-M)(M+P_0T_0)}]^2}.
\end{equation}

\subsection{Optimal Antenna Size}

Lastly, we consider the optimal size of the antenna array for the system with the proposed position-aided estimation scheme. It is assumed that the number of receive antennas is always equal to that of transmit antennas. Since the total number of transmit antennas is $M=M_{e}\cdot M_{g}$ and the value of $M_e$ is given by (\ref{eqn:maximum number of antenna}), it remains to determine the value of $M_{g}$. We consider this problem from the viewpoint of maximizing the multiplexing gain of the system, namely the degrees-of-freedom (DoF) of the system, which is defined as follows \cite{Zheng_25}
\begin{equation}\label{eqn:definition of dof}
  \text{DoF}=\lim \limits_{P_0 \rightarrow \infty} \frac{R_{\text{L}}}{\log_2 P_0}
\end{equation}

\emph{\textbf{Proposition 3:} For a training-based system with position-aided channel estimation, the optimal number of transmit antennas} $M^*$ \emph{in terms of maximizing DoF is}
\begin{equation}\label{eqn:optimization for proposed system}
M^*=\frac{T_0}{2} \cdot M_e, 
\end{equation}
i.e., $M_g^*=\frac{T_0}{2}$ and $M_e$ is given by (\ref{eqn:maximum number of antenna}).
\begin{IEEEproof}
 See the Appendix D.
\end{IEEEproof}

Similar to \emph{Proposition 3}, we can get the optimal number of transmit antennas for the conventional training system by setting $M_e=1$ and $M_g=M$, which is summarized as follows.

\emph{\textbf{Corollary 2:} For the training-based system with conventional training, the optimal number of transmit antennas in terms of maximizing DoF is}
\begin{equation}\label{eqn:optimal throughput in traditional system}
  M^*=\frac{T_0}{2}.
\end{equation}

\section{Simulation Results}

\subsection{Comparison Between Two Training Schemes}

We first compare the throughput performance of the proposed position-aided channel estimator to that of the conventional one, the explicit expressions of which are given in (\ref{eqn:optimal capacity under proposed system in the new system}) and (\ref{eqn:optimal capacity in conventional system}), respectively. For the fairness of comparison, we assume that the antenna sizes under the two estimation schemes are the same in this subsection.

\begin{figure}[!t]
\centering
\includegraphics[width=3.3 in]{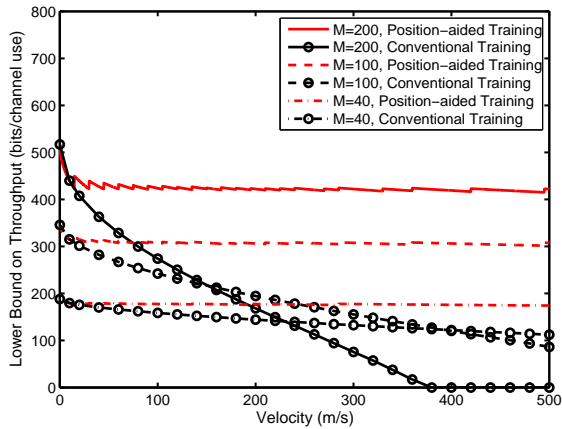}
\caption{Throughput comparison between the proposed position-aided channel estimator and the conventional estimator as a function of velocity $v_0$ when the average SNR is $30$dB.}
\label{fig:comparison between two schemes}
\end{figure}

Specifically, it is assumed that the carrier wavelength $\lambda_0=0.15$m (i.e., the carrier frequency is $2$GHz), the bandwidth $B_0=10$MHz, the coherence time of environment $t_0=5$ms, and the length of each signal block is equal to the twentieth of the coherence time of channel ($\xi_0=20$). The average SNR value is $30$dB. We consider the case that the number of receive antennas is the same as that of transmit antennas. Fig. \ref{fig:comparison between two schemes} plots the throughput of a training-based system under the proposed position-aided channel estimator and the conventional estimator as a function of the velocity $v_0$ when the number of antennas $M$ is $100$ and $200$, respectively. In the system with position-aided channel estimation, the value of $M_e$ is given by (\ref{eqn:maximum number of antenna}), and the corresponding $M_g$ is equal to $\lceil \frac{M}{M_e}\rceil$; if the value of $M$ is not a multiple of $M_e$, some extra zero columns can be added to the end of the last group $\bm{Q}_k^{M_{g}}$ to match with the formulation in (\ref{eqn:optimal capacity under proposed system in the new system}).

It is seen from Fig. \ref{fig:comparison between two schemes} that the throughput of the conventional training scheme deteriorates significantly as the relative velocity increases, especially when the antenna size is large. This is because the training phase occupies too many channel uses. In particular, the throughput can even become zero when the velocity is large enough, which highlights the main motivation of this work to propose the concept of position-aided training: to reduce the estimation overhead in highly mobile environments. In contrast, the performance of the position-aided estimation scheme deteriorates a little and is nearly independent of the velocity due to the exploitation of the spatio-temporal correlation in a mobile environment. A higher mobility leads to smaller value of $T_0$, which reduces the duration of data phase. However, it also makes it possible to group more columns together to share the common training signal based on (\ref{eqn:maximum number of antenna}), which can reduce the portion of training phase in a block. As a result, the system with position-aided channel estimation can achieve a robust performance with respect to mobility, even when $M$ is just on the order of tens (such as $M=40$). Significant improvement can be achieved if we employ position-aided channel estimation for the large-scale MIMO system in the high-speed railway scenarios.

\subsection{Performance Comparison under the Optimal Antenna Array Size}

\begin{figure}[!t]
\centering
\includegraphics[width=3.3 in]{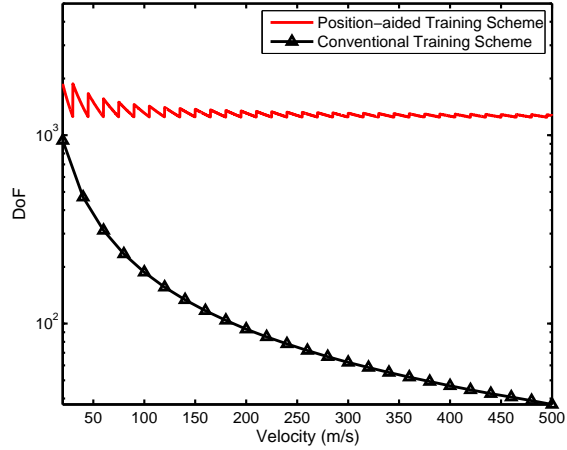}
\caption{The DoF of the training-based system with position-aided training and conventional training as a function of velocity $v_0$, when the optimal antenna sizes in (\ref{eqn:optimization for proposed system}) and (\ref{eqn:optimal throughput in traditional system}) are adopted.}
\label{fig:simulation results three in group system}
\end{figure}

In this subsection, we compare the performance of the two channel estimation schemes under their respective optimal antenna sizes which are given by (\ref{eqn:optimization for proposed system}) and (\ref{eqn:optimal throughput in traditional system}). Fig.~\ref{fig:simulation results three in group system} depicts the DoF performance of the training-based system as a function of velocity $v_0$ with the position-aided training scheme and the conventional training scheme. It can be observed that the DoF with conventional training decreases with the velocity $v_0$. The DoF with position-aided channel estimation is significantly higher than that of the conventional one, especially when the speed is high, which is consistent with the results in Fig. \ref{fig:comparison between two schemes} for the case of fixed antenna size. It should be noted that the discontinuity phenomenon in the performance curves is caused by the round-off operation in calculating $M_e$ and $M_g$.

%

\begin{figure}[!t]
\centering
\subfloat[]{
\label{fig:simulation result in new system}
\begin{minipage}[t]{0.5\textwidth}
\centering
\includegraphics[width=3.2 in]{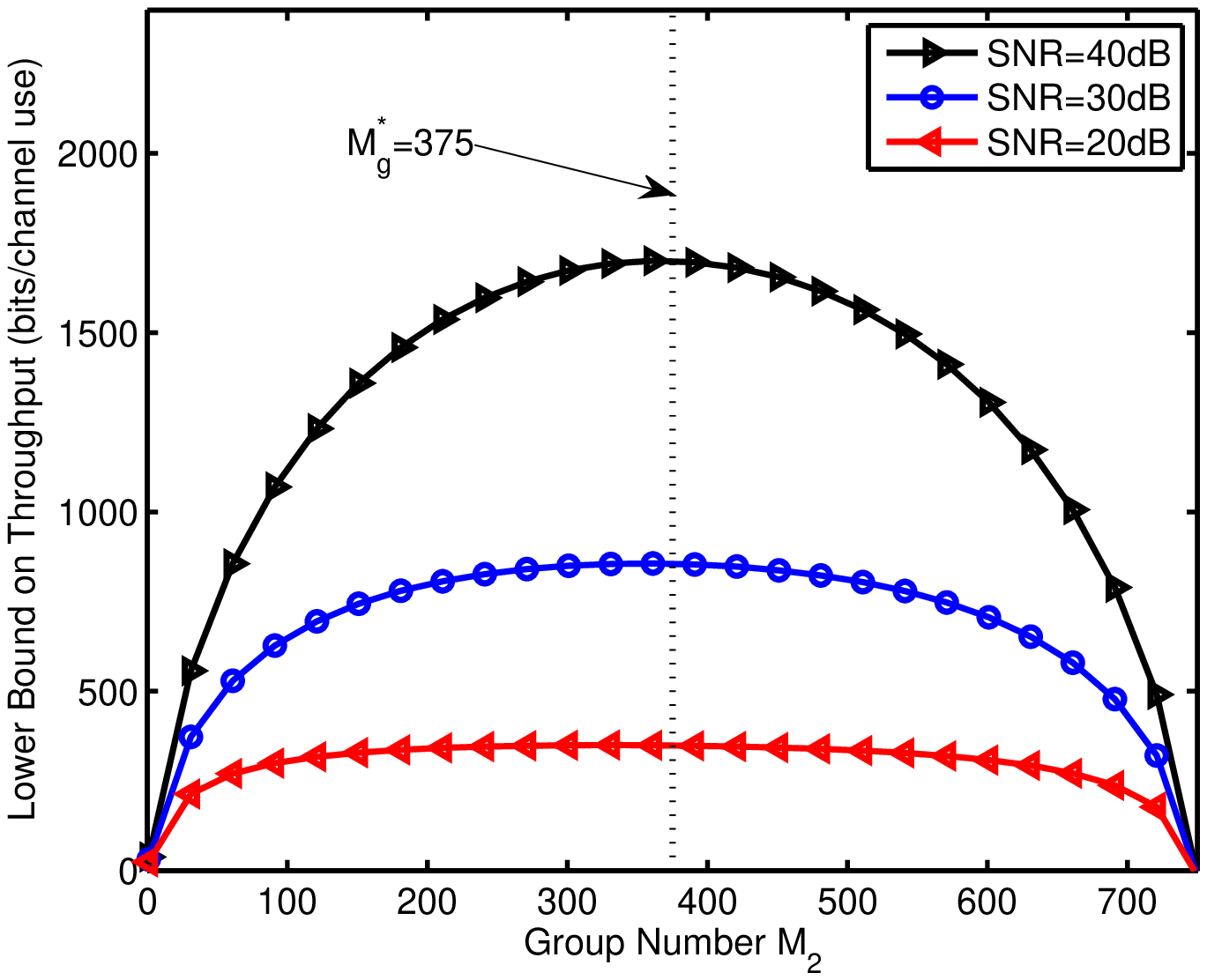}
\end{minipage}
}
\subfloat[]{
\label{fig:simulation result in conventional system}
\begin{minipage}[t]{0.5\textwidth}
\centering
\includegraphics[width=3.2 in]{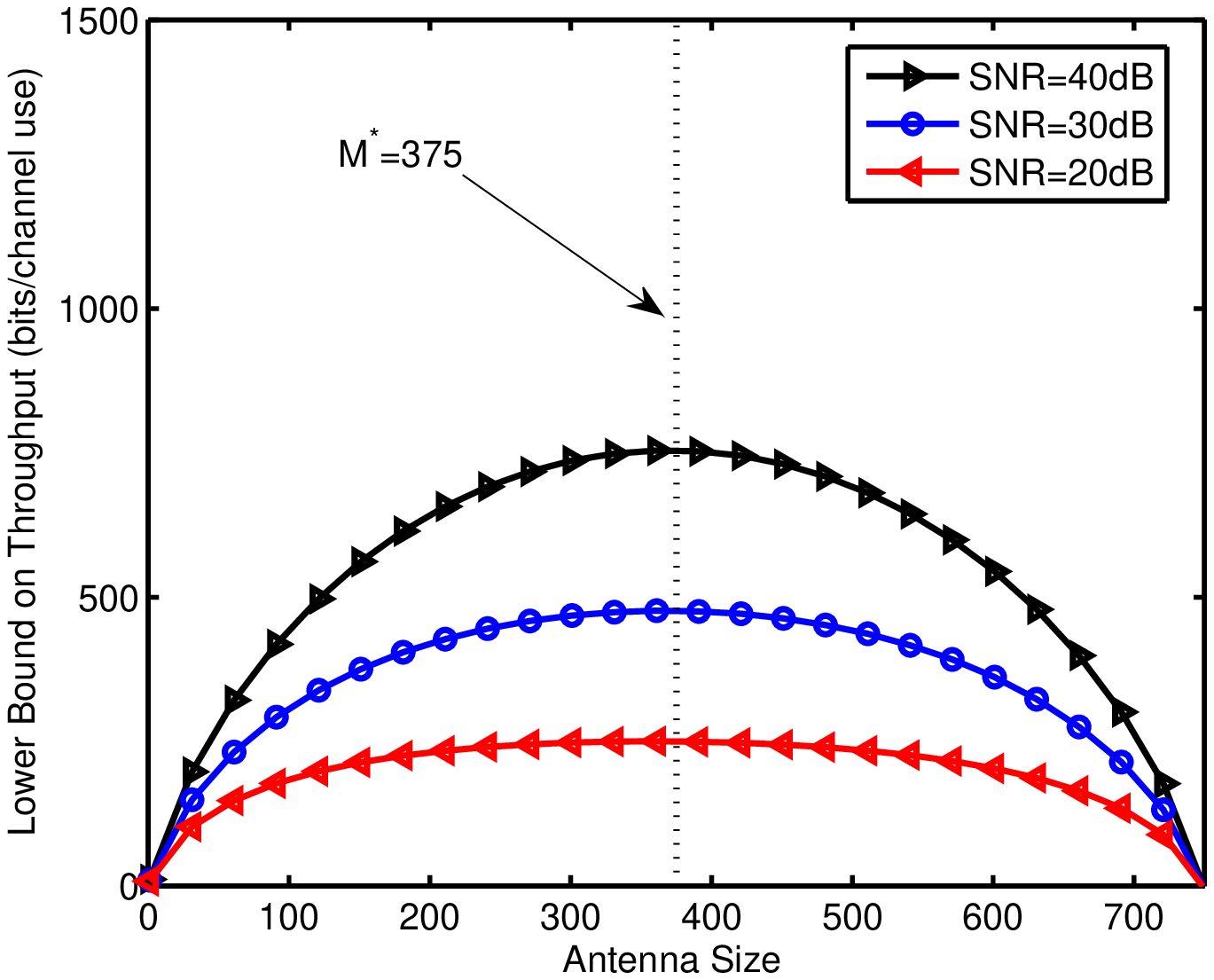}
\end{minipage}
}
\caption{The system throughput as a function of antenna size $M$ when $v_0=100$m/s: (a) with position-aided channel estimation scheme, (b) with conventional training scheme.}
\end{figure}

The optimal antenna size $M^*$ in (\ref{eqn:optimization for proposed system}) is obtained based on DoF maximization in the high SNR regime. Let us examine the optimality based on numerical simulation when the SNR is not so high. Assuming that the velocity $v_0=100$m/s and the other parameters are just the same as those in the previous subsection, Fig. \ref{fig:simulation result in new system} depicts the system throughput with position-aided channel estimation as a function of group number $M_g$ under different average SNR values. The ideal optimum group number calculated by (\ref{eqn:optimization for proposed system}) is $M_g^*=375$ as displayed in Fig.~\ref{fig:simulation result in new system}. It can be observed that the practical optimal value for $M_g$ is very close to $375$ even when the SNR is only $20$dB. Likewise, Fig. \ref{fig:simulation result in conventional system} depicts the throughput under conventional training scheme as a function of antenna size $M$ when $v_0=100$m/s. Similar results can be observed from it.

%

\subsection{Performance with $L_0>1$}

\begin{figure}[!t]
\centering
\includegraphics[width=3.3 in]{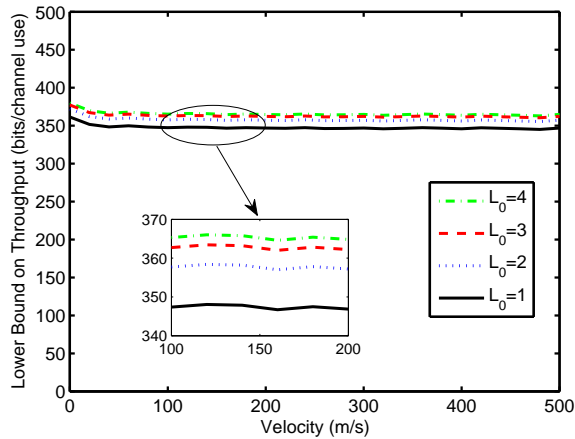}
\caption{The system throughput with position-aided channel estimation as a function of velocity for $L_0=1,2,3,4$ and $SNR=20$dB.}
\label{fig:throughput_vs_velocity_in_general_system}
\end{figure}

\begin{figure}[!t]
\centering
\includegraphics[width=3.3 in]{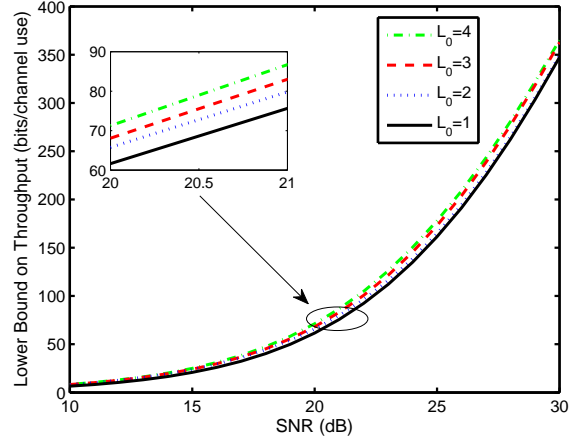}
\caption{The system throughput with position-aided channel estimation as a function of SNR for $L_0=1,2,3,4$ and $v_0=100$m/s.}
\label{fig:throughput_vs_snr_in_general_system}
\end{figure}

The analysis in Section IV and the above numerical results concentrate on the case of $L_0=1$ in (\ref{eqn:MMSE estimator for the first column in each group}) and (\ref{eqn:conditional mean estimator for j column}). We now consider the general case with $L_0>1$ via simulations.
It is assumed that the system parameters are the same as those in Section V.A. The antenna size at the transmitter and receiver are $M=N=200$. The value of $M_e$ is given by (\ref{eqn:maximum number of antenna}) and the corresponding $M_g$ is set as $\lceil \frac{M}{M_e}\rceil$. Fig. \ref{fig:throughput_vs_velocity_in_general_system} plots the system throughput with the position-aided channel estimation as a function of the velocity with $L_0=1,2,3,4$ and $SNR=20$dB. The training interval in both cases is set as $T_\tau=M_g$. Besides, for a fair comparison, the uniform power distribution is adopted, i.e., $P_\tau=P_d=P_0$.
Fig. \ref{fig:throughput_vs_snr_in_general_system} plots the system throughput under position-aided channel estimation scheme as a function of SNR when the velocity is $100$m/s, $M=N=200$ and $L_0=1,2,3,4$. From Figs. \ref{fig:throughput_vs_velocity_in_general_system} and \ref{fig:throughput_vs_snr_in_general_system}, it is seen that there is only a slight gain with $L_0>1$ compared with $L_0=1$. Thus, we strongly recommend to employ the case with $L_0=1$ in a highly mobile large-scale MIMO system, to achieve a considerable improvement with low complexity.

\section{Conclusions}

We have proposed a position-aided channel estimation scheme for training-based large-scale MIMO systems to reduce the pilot overhead in high-speed railway communications. In this concept, only a subset of the transmit antennas need to send pilot symbols during the training phase of each block.
The entire channel matrix can be estimated from the initial estimate of the submatrix with the help of position information by exploiting the spatio-temporal correlation structure of the channel.
We have also developed a framework of optimizing the training interval, power allocation and antenna size for the proposed position-aided training system. A salient feature of the proposed scheme is that the system throughput remains invariant as the transmitter's moving speed varies, whereas for the system that employs conventional training, the throughput deteriorates rapidly as the speed increases and even becomes zero with very high mobility.


\section*{Appendix A: Proof of Proposition 1}

The condition $\Gamma(\eta_1,\eta_1',\eta_1'')>1$ is equivalent to
\begin{equation}\label{eqn:equivalent inequility}
  \frac{{\eta_1'}^2+{\eta_1''}^2-2\eta_1\eta_1'\eta_1''\sigma_0^2}{1-\eta_1^2\sigma_0^4} > 1.
\end{equation}

Since both $\eta_1$ and $\sigma_0^2$ belong to $(0,1)$, we have $1-\eta_1^2\sigma_0^4>0$. By some manipulations, (\ref{eqn:equivalent inequility}) is equivalent to
\begin{align}\label{eqn:equivalent inequility 2}
  &  \eta_1^2\sigma_0^4-2\eta_1\eta_1'\eta_1''\sigma_0^2+({\eta_1'}^2+{\eta_1''}^2-1) > 0\\
  \Rightarrow \,\,&  \sigma_0^2 \in \Big(0, \frac{2\eta_1\eta_1'\eta_1''-\sqrt{4\eta_1^2{\eta_1'}^2{\eta_2''}^2-4\eta_1^2({\eta_1'}^2 +{\eta_1''}^2-1)}}{2\eta_1^2}\Big).
\end{align}

From (\ref{eqn:the expression for eta_i' and i''}), it can be seen that $\eta_1'$ and $\eta_1''$ are  functions of $z_{k_0}^{i_0}$, $z_{k}^{j_0}$ and $ z_{k_0+1}^{i_0}$, the values of which are different for different columns. In order to guarantee that $\Gamma(\eta_1,\eta_1',\eta_1'')>1$, we then need $\sigma_0^2\in (0,\Omega)$, where
\begin{equation}\label{eqn:expression for omega}
  \Omega=\min \limits_{ \{z_{k}^{j_0},j_0=1,2,...M\} }
  \frac{2\eta_1\eta_1'\eta_1''-\sqrt{4\eta_1^2{\eta_1'}^2{\eta_2''}^2-4\eta_1^2({\eta_1'}^2+ {\eta_1''}^2-1)}}{2\eta_1^2}.
\end{equation}

Thus, $\Gamma(\eta_1,\eta_1',\eta_1'')>1$, if the SNR value during the training phase meets the following constraint on the condition
\begin{equation}\label{eqn:snr in training phase}
  \frac{P_\tau T_\tau}{M_g}<\frac{\Omega}{1-\Omega}.
\end{equation}

\section*{Appendix B: Proof of Lemma 1}

As observed from (\ref{eqn:transmission performance of new system})--(\ref{eqn:expression of effective snr}), the power allocation strategy $\{P_\tau,P_d\}$ only affects the throughput via $\rho_{\text{eff}}$. Thus, maximizing $\rho_{\text{eff}}$ with respect to $(P_\tau, P_d)$ is equivalent to maximizing $R_{\text{L}}$. We use a similar formulation as that in \cite{Hassibi_5}. That is, letting $\alpha$ be the fraction of total transmit energy that is dedicated to data phase, we have
\begin{equation}\label{eqn:fraction of transmit power for data phase and training phase}
  P_d T_d=\alpha P_0T_0,\,\, P_\tau T_\tau=(1-\alpha)P_0T_0,  \,\, 0<\alpha<1.
\end{equation}

Then, we can rewrite the effective SNR in (\ref{eqn:expression of effective snr}) as
\begin{equation}\label{eqn:rewritten expression of effective snr}
  \rho_{\text{eff}}=\frac{P_d \frac{P_\tau}{M_g} T_\tau}{1+P_d+\frac{P_\tau}{M_g} T_\tau}
  =\frac{\alpha \frac{P_0T_0}{M_gT_d} (1-\alpha)P_0T_0}
  {1+\alpha \frac{P_0T_0}{T_d}+(1-\alpha)\frac{P_0T_0}{M_g}}
  =\frac{P_0^2T_0^2}{\frac{M_g(T_d+P_0T_0)}{1-\alpha}+\frac{T_d(M_g+P_0T_0)}{\alpha}}.
\end{equation}

Denote $\mathfrak{L}(\alpha)=\frac{M_g(T_d+P_0T_0)}{1-\alpha}+\frac{T_d(M_g+P_0T_0)}{\alpha}$. As a result, minimizing $\mathfrak{L}(\alpha)$ is equivalent to maximizing $\rho_{\text{eff}}$. We have
\begin{align}\label{eqn:first order derivative}
  & \frac{\partial\mathfrak{L}}{\partial \alpha}=\frac{M_g(T_d+P_0T_0)}{(1-\alpha)^2}-\frac{T_d(M_g+P_0T_0)}{\alpha^2}, \\
  & \frac{\partial^2\mathfrak{L}}{\partial \alpha^2}=2\cdot\frac{M_g(T_d+P_0T_0)}{(1-\alpha)^3}+2\cdot\frac{T_d(M_g+P_0T_0)}{\alpha^3}.
\end{align}

Since $\frac{\partial^2\mathfrak{L}}{\partial \alpha^2}>0$, $\mathfrak{L}(\alpha)$ is convex and has an unique minimum, which is
\begin{equation}\label{eqn:minimal value of function}
  \mathfrak{L}_{\text{min}}=\big[\sqrt{M_g(T_d+P_0T_0)}+\sqrt{T_d(M_g+P_0T_0)}\big]^2.
\end{equation}
By solving $\frac{\partial\mathfrak{L}}{\partial \alpha}=0$, we obtain
\begin{equation}\label{eqn:correponding value of alpha}
  \alpha^*=\frac{\sqrt{T_d(M_g+P_0T_0)}}{\sqrt{M_g(T_d+P_0T_0)}+\sqrt{T_d(M_g+P_0T_0})}.
\end{equation}

\section*{Appendix C: Proof of Lemma 2}

Let us consider the monotonicity of the throughput function $R_{\text{L}}$ in (\ref{eqn:transmission performance of new system}) with respect to the variable $T_d$ under the optimal power allocation presented in \emph{Lemma 1}. Plugging (\ref{eqn:power allocation coefficient in the new system}) into (\ref{eqn:expression of effective snr}), the effective SNR with optimal power allocation can be rewritten as
\begin{equation}\label{eqn:expression of effective snr in lemma 4.3}
  \rho_{\text{eff}}=\frac{P_0^2T_0^2}{[\sqrt{M_2(T_d+P_0T_0)}+\sqrt{T_d(M_2+P_0T_0)}]^2}.
\end{equation}

Assuming $\lambda_i$ is the \mbox{$i$-th} nonnegative singular value of the matrix $\frac{ \overline{H}_k \overline{H}_k^*} {M}$, the throughput function can be expressed as
\begin{equation}\label{eqn:rewritten the throughput function}
  R=\sum \limits_{i} \mathbb{E} \Big\{ \frac{T_d}{T_0} \cdot  \ln \big(1+\rho_{\text{eff}} \lambda_i \big) \Big\},
\end{equation}
where we use the natural logarithm to instead of $\log_2$ for convenience, and the expectation operation is over $\lambda_i$.

Let $R_i(T_d)$ be $\mathbb{E} \big\{ \frac{T_d}{T_0} \cdot  \ln (1+\rho_{\text{eff}} \lambda_i) \big\}$. The first order derivative of $R_i(T_d)$ with respect to $T_d$ is
\begin{equation}\label{eqn:first order derivative in lemma 4.3}
  \frac{\partial R_i(T_d)}{\partial T_d}=\mathbb{E} \Big\{ \frac{1}{T_0}\log_2(1+\rho_{\text{eff}}\lambda_i)+ \frac{T_d}{T_0} \frac{\lambda_i}{1+\rho_{\text{eff}}\lambda_i} \frac{\partial \rho_{\text{eff}}}{\partial T_d} \Big \},
\end{equation}
where
\begin{equation}\label{eqn:first order derivative for effective snr in lemma 4.3}
\begin{split}
  \frac{\partial \rho_{\text{eff}}}{\partial T_d}= & \,- \tfrac{P_0^2T_0^2 \cdot [2\sqrt{M_g(T_d+P_0T_0)}+2\sqrt{T_d(M_g+P_0T_0)}]}{\{ [\sqrt{M_g(T_d+P_0T_0)}+\sqrt{T_d(M_g+P_0T_0)}]^2\}^2} \cdot \big[\tfrac{\sqrt{M_g}}{2\sqrt{T_d+P_0T_0}}+ \tfrac{1}{2}\tfrac{\sqrt{M_g+P_0T_0}} {\sqrt{T_d}} \big] \\
  =&\,  \frac{1}{T_d} \cdot \tfrac{P_0^2T_0^2}{[\sqrt{M_g(T_d+P_0T_0)}+\sqrt{T_d(M_g+P_0T_0)}]^2} \\
  &\, \cdot \tfrac{\big[\sqrt{M_g(T_d+P_0T_0)}+\sqrt{T_d(M_g+P_0T_0)}\big]\cdot \big[\tfrac{T_d\sqrt{M_g}}{\sqrt{T_d+P_0T_0}}+\sqrt{T_d(M_g+P_0T_0)} \big]}{[\sqrt{M_g(T_d+P_0T_0)}+\sqrt{T_d(M_g+P_0T_0)}]^2}.
\end{split}
\end{equation}
Because
\begin{equation}\label{eqn:some manipulation 2 in lemma 4.3}
  \tfrac{T_d\sqrt{M_g}}{\sqrt{T_d+P_0T_0}}<\sqrt{M_g(T_d+P_0T_0)},
\end{equation}
we have
\begin{equation}\label{eqn:some manipulation 3 in lemma 4.3}
\begin{split}
  & \big[\sqrt{M_g(T_d+P_0T_0)}+\sqrt{T_d(M_g+P_0T_0)}\big]\cdot \big[\frac{T_d\sqrt{M_g}}{\sqrt{T_d+P_0T_0}}+\sqrt{T_d(M_g+P_0T_0)} \big] \\
  <& \big[\sqrt{M_g(T_d+P_0T_0)}+\sqrt{T_d(M_g+P_0T_0)}\big]^2.
\end{split}
\end{equation}
Substituting (\ref{eqn:some manipulation 3 in lemma 4.3}) into (\ref{eqn:first order derivative for effective snr in lemma 4.3}), we can get
\begin{equation}\label{eqn:some manipulation 4 in lemma 4.3}
  -\frac{T_d \partial \rho_{\text{eff}}}{\partial T_d}< \rho_{\text{eff}}.
\end{equation}

Besides, the function $\ln(1+x)-x/(1+x)\geq 0$ for all $x \geq 0$, since it is zero at $x=0$ and an increasing function for $x \geq 0$. Thus, combining the results in (\ref{eqn:first order derivative in lemma 4.3}) and (\ref{eqn:some manipulation 4 in lemma 4.3}), we can get
\begin{equation}\label{eqn:some manipulation 5 in lemma 4.3}
  \frac{1}{T_0}\Big[\ln (1+\rho_{\text{eff}}\lambda_i)- \frac{\rho_{\text{eff}}\lambda_i}{1+\rho_{\text{eff}}\lambda_i}\Big] \geq 0.
\end{equation}

Thus,
\begin{equation}\label{eqn:some manipulation 6 in lemma 4.3}
  \frac{\partial R_i(T_d)}{\partial T_d}=\mathbb{E} \Big\{ \frac{1}{T_0}\log_2(1+\rho_{\text{eff}}\lambda_i)+ \frac{T_d}{T_0} \frac{\lambda_i}{1+\rho_{\text{eff}}\lambda_i} \frac{\partial \rho_{\text{eff}}}{\partial T_d} \Big \}\geq 0.
\end{equation}

In summary, based on (\ref{eqn:some manipulation 6 in lemma 4.3}), $R_i(T_d)$ is a monotonically increasing function with respect to $T_d$ for arbitrary $\lambda_i$. Thus, the throughput function $R_{\text{L}}$ in (\ref{eqn:rewritten the throughput function}) is a monotonically increasing function with respect to $T_d$. To get better performance, $T_d$ should be as large as possible. Thus, the optimal training interval is equal to the number of group $M_g$ in the proposed position-aided group training scheme, which is the minimum value that is required for learning the matrix $\bm{G}_k$.

\section*{Appendix D: Proof of Proposition 3}

Using (\ref{eqn:optimal capacity under proposed system in the new system}) and (\ref{eqn:effective snr value in new system}), and assuming that $M=N$, we can get
\begin{equation}\label{eqn:expression of throughput}
\begin{split}
  R_{\text{\tiny{L}}}\,&=\,  \frac{T_0-M_g}{T_0} \cdot \mathbb{E} \Big\{ \log_2 \,\det \big(\bm{I}_N+\rho_{\text{eff}}^{*} \frac{ \overline{\bm{H}}_k\overline{\bm{H}}_k^H}{M} \big) \Big\} \\
  &=\,\frac{T_0-M_g}{T_0} \cdot \sum_{i=1}^{M_e \cdot M_g} \mathbb{E} \Big\{ \log_2 \, \big(1+P_0 \cdot \tfrac{P_0T_0^2}{[\sqrt{M_g(T_0-M_g+P_0T_0)}+\sqrt{(T_0-M_g)(M_g+P_0T_0)}]^2} \cdot \lambda_i^2 \big) \Big\},
\end{split}
\end{equation}
where $\lambda_i^2$ denotes the \mbox{$i$-th} singular value of $\frac{ \overline{\bm{H}}_k\overline{\bm{H}}_k^H}{M}$.

Hence, we have
\begin{equation}\label{eqn:an approximation in high SNR regime}
\begin{split}
  \lim \limits_{P_0\rightarrow \infty} \,\,\,R_{\text{L}}=&
   \frac{T_0-M_g}{T_0} \cdot M_e \cdot M_g \big[\log_2 P_0 +o(\log_2 P_0 )\big] \\
   =& \Big\{-\big(M_g-\frac{T_0}{2}\big)^2+\frac{T_0^2}{4} \Big\} \cdot \frac{ M_e \cdot [\log_2 P_0+o(\log_2 P_0)]}{T_0}.
\end{split}
\end{equation}
where $o(x)$ is defined as $\lim \limits_{x\rightarrow 0}\frac{o(x)}{x}=0$.

Then, the degrees-of-freedom metric becomes
\begin{equation}\label{eqn:definition for DoF}
  \text{DoF}=\lim \limits_{P_0 \rightarrow \infty} \frac{R_{\text{L}}}{\log_2 P_0} =\Big\{-\big(M_g-\frac{T_0}{2}\big)^2+\frac{T_0^2}{4} \Big\} \cdot \frac{ M_e}{T_0}.
\end{equation}
which is maximized by $M_g^*=\frac{T_0}{2}$.
\ifCLASSOPTIONcaptionsoff
  \newpage
\fi

\end{document}